\documentclass[floatfix,showpacs,showkeys,prb,twocolumn]{revtex4}
\usepackage{epsfig}
\usepackage{graphicx}
\usepackage{amssymb}
\usepackage{amsmath}
\usepackage{hyperref}

\usepackage{color}

\begin{document}

\title{Magnetic Neutron Scattering in Solid Oxygen and its Applications to Ultracold Neutron Production}
\date{\today}

\author{C.-Y. Liu}
\email{cl21@indiana.edu}
\author{C.M. Lavelle}
%\author{D. Salvat}
\affiliation{Physics Department, Indiana University, Bloomington, IN 47408}

\author{C.M. Brown}
\affiliation{NIST Center for Neutron Research, National Institute of Standards and Technology, Gaithersburg, Maryland 20899}

\begin{abstract}
%We present neutron scattering data on oxygen solids at the saturated vapor pressure. 
Using neutron scattering, we investigate the static and dynamic structure functions $S(Q,\omega)$ of 3 distinct solid phases (using powder average techniques), and characterize the corresponding modes of energy excitation, focusing primarily on the dynamics of spin interaction. 
With the $S(Q,\omega)$ maps, we extract the temperature dependent production rate and upscattering cross section, that are relevant for understanding experimental data on ultracold neutron production in solid oxygen at the saturated vapor pressure. 
\end{abstract}

\pacs{29.25.Dz, 28.20.Gd, 28.20.-v}
\keywords{Ultra-cold Neutron; UCN; Neutron moderation; Solid Oxygen; Magnetic Scattering; Magnon}

\maketitle

\section{Introduction}

Because of its non-zero electronic spin ($S$=1), oxygen displays a uniquely rich phase diagram when compared to other cryocrystals made of diatomic molecules, such as N$_2$, H$_2$ and CO. 
%The low solidification temperature of oxygen is a result of the weak vander Waal force. 
At the saturated vapor pressure, there exist 3 distinct solid phases: $\gamma$ (44 $\sim$ 55~K, cubic lattice, paramagnetic), $\beta$ (24 $\sim$ 44~K, rhombohedral lattice, short range anti-ferromagnetic(AF)), and $\alpha$ phase ( $<$ 24~K, monoclinic lattice, 2-D AF)~\cite{Freiman04}. 
When subject to high pressures, oxygen loses it magnetism and this Mott insulator becomes metallic. Upon further pressurization, the solid becomes superconducting at temperatures below 0.2~K. 
While there exists great theoretical interest with the complexity of these phase transitions of this seemingly simple system, in the paper, we will focus on gaining understanding of the spin dynamics in the low temperature phases at low pressures, and illustrate their potential applications of ultracold neutron (UCN) production.

The magnetic excitations in solid oxygen (s-O$_2$) bring about an opportunity to enhance the superthermal UCN production through inelastic magnetic scattering, beyond the typical phonon couplings already applied in superfluid helium and solid deuterium with a great success. 
In a superthermal UCN source,
the dominant upscattering loss is suppressed by reducing the number of thermally populated energy quanta (phonon or magnon) through cooling of the source material, whereas the production is relatively temperature-independent as long as the inelastic scattering channels remain open and unperturbed. The phase space of the neutrons can be compressed through energy dissipation by exciting the collective modes in the solid.
In s-O$_2$, the magnetic scattering length is comparable to that of nuclear 
scattering, and its spin degree of freedom allows additional inelastic neutron scattering through magnetic coupling that has never been applied to neutron moderations.
In this paper, we report results on modern inelastic neutron scattering (INS) in attempt to understand the physics of magnetic excitations in s-O$_2$ relevant to our UCN production experiment. 

%In $\alpha$ phase, our high resolution INS data confirm the 2D AF magnon model used in the work by Stephens and coworkers~\cite{Stephens86}, however, we also report new values of anisotropy energies that prefer the 2D Ising model than the XY model suggested in the paper~\cite{Stephens86} that led to our previous over-estimate of the UCN production.
%With the corrected coupling constants, we attain a better understanding of the temperature dependence of the excitation energies. In the $\beta$ phase, we have reported an unexpected large UCN production. The INS analysis shows that the enhanced UCN production is a result of a dispersionless magnetic excitation in this geometrically frustrated spin system. These excitation modes were previously~\cite{Chahid1993} characterized using a phenomenological model of coupled paramagnets, however, the ground state configuration must be highly unstable and the INS data shows many distinct features that warrant further theoretical investigations. 

%\section{Background}

At low temperatures, the spin ordered state of s-O$_2$ is a Neel state with 
a AF ordering in the basal plane. As the temperature rises, the spin-spin correlation remains strong enough to evade thermal fluctuations, resulting in a short-range spin order in the $\beta$ phase, and the spin correlation persists in the $\gamma$ and liquid phases up to relatively high temperatures. 
The spin system does not follow the simple order-disorder transition, but instead an intermediate phase exists between the spin-ordered $\alpha$ phase and spin-disordered $\gamma$ phase. This intermediate phase is best categorized as a geometrically frustrated spin system, a topic of which much work are actively ongoing to formulate viable wave-functions to calculate energy excitations~\cite{Lee86, Coldea04}. 
Traditionally, the spin order in the $\beta$ phase was described by a helimagnet or a three-sublattice Yaffet-Kittel structure~\cite{Meier85, Jansen86}.
However, due to its low spin number, the system should really be treated quantum mechanically with discrete spin quantizations. The classical spin models might not produce the correct excitation spectrum~\cite{Jansen87a}, especially in the  
strongly coupled, geometrically frustrated $\beta$ phase. 
%is stable well above room temperature when pressurized.
At low temperatures under zero pressure, the frustration is relieved by a structural distortion as the hexagonal lattice structure is slightly sheared into a monoclinic structure with reduced symmetry.

The monoclinic unit cells of the $\alpha$ and $\beta$ phases can be transformed into each other by a small inhomogeneous deformation of the closely packed basal planes, driven by the spin ordering. 
%monoclinic distortion of the regular hexagon at the $\beta-\alpha$ transition.
The $c_{\perp}$ lattice constant increases when approaching the phase transition from high temperature, leading to the dominant 2D spin alignment, with 
AF exchange interaction and an easy-axis anisotropy. 
%The magnetic interaction prefers monoclinic deformation, and thus the hexagonal lattice is stablized by the elastic forces. 
The two-sublattice AF monoclinic lattice transforms discontinuously into the three-sublattice AF hexagonal lattice and undergoes a second order phase transition into the magnetically disordered structure (albeit with short range spin correlations).
It was suggested that magneto-elastic coupling changes the order of the transition from the second to the first one with a sizable latent heat~\cite{daSilva95}. 
In addition, it was not yet understood why the $\alpha-\beta$ transition happens at a temperature as low as 24~K, despite of its strong exchange field. 
Inter-plane exchange, spin-orbital, and magneto-dipole interactions all have the opposite effect to cut down fluctuations at the small-$k$. 
This can probably be resolved if the quantum fluctuations in the quasi-two-dimensional Heisenberg AFM are taken into account.

Concerning the UCN production, we are most interested in the energy excitation modes that incident cold neutrons (CN) can efficiently couple to. 
While the excitation in $\alpha$-O$_2$ can be described by magnons using spin wave theory, the situation in $\beta$-O$-2$ still needs to be clarified.
The first modern many-body calculation of excitations in s-O$_2$ revealed interesting behaviors~\cite{daSilva95}. In the $\beta$ phase, the ground state has a two-fold degeneracy with a low lying first excited state at 0.8~meV. Because this excited state corresponds to the center of the Brillouin zone, as opposed to the zone boundary as with typical magnon excitations, the authors speculated the existence of bound pairs of spin waves in $\beta$ phase. This hints to the formation of a spin singlet solid, a necessary condition to attain the spin liquid state. 
Another Monte-Carlo simulation which allows a deformable cell~\cite{LeSar88} 
also showed an abrupt increase in spin dynamics in the $\beta$ phase beyond that of the stable $\alpha$ phase. 
%On the other hand, a molecular dynamic calculation which incorporates the spin degrees of freedom~\cite{Jansen87a} pointed out the importance of the spin-orbital coupling. In the molecular ground state, the electronic spin of the valence electrons points perpendicular to the molecular axis. Therefore, any molecular libration would result in the tilt motions of spin.      
Comparing the excitation spectrum in s-O$_2$ to other well studied frustrated spin systems,
such as $\alpha$-NaMnO$_2$ (with $S=2$)~\cite{Stock09} and Cs$_2$CuCl$_4$ ($S=1/2$)~\cite{Sato03}, we find many similarities in the INS data. 

\section{Experiments}
We measured the total cross section of cold neutrons using a standard transmission measurement, as a part of the UCN production experiment at Lujan Center neutron flight path 12 at the Los Alamos Neutron Science Center (LANSCE). Details of the experimental setup are described in ~\cite{Lavelle2010}. 
The INS experiment was carried out using the Disk Chopper Spectrometer (DCS)~\cite{Copley03} at the National Institute of Standards and Technology (NIST) Center for Neutron Research (NCNR), where neutrons are produced from a 20 MW fission reactor. 
%A series of pulsing and monochromating choppers prepare incident neutron beam on DCS.
Most of the data were taken using neutrons at 2.3 $\mbox{\AA}$ (15.4~meV). With this wavelength, the energy resolution (FWHM) of the instrument is 1 meV at elastic scattering, and improves as the neutron time-of-flight (TOF) increases to a resolution of $\sim$ 500 $\mu$eV at energy transfer $\omega = 15$~meV. 
The target cell was mounted inside an ``orange" helium-flow cryostat. 
Liquid oxygen was first condensed from gaseous O$_2$ (at 99.999\%) in an annular cell made of aluminum, with an outer diameter of 12.5 mm and an inner diameter of 10.5 mm, leaving an annular thickness of 2 mm. With the height of 110 mm, the cell has a total volume of 6.1 cm$^3$. 
The beam was collimated to 90 mm of total height, and illuminated the entire width of the target cell.
Inside the cell, the liquid was subsequently solidified and cooled under the saturated vapor pressure. 
A s-O$_2$ sample in the polycrystalline form was attained under typical cooling procedures, as the thermal contraction, in particular the abrupt 5\% change of the molar volume through the $\gamma$-$\beta$ transition, would most likely crack the bulk solid into many micro-crystals. 
The cryostat was allowed to stabilize at each set temperature for 
1 hour before beginning a measurement.  
%Data were taken at 29 temperatures from 4-47 K were recorded at about 4 hours each.
Data were collected at 2~K temperature step throughout the $\beta$ and $\alpha$ phase with smaller temperature steps around the $\alpha$-$\beta$ transition.
%In the final day of the 6-day beam time, 
In the final step, the cell was warmed up to totally evacuate oxygen and the background data were collected with the empty aluminum cell at 24~K. Calibration data on detector efficiencies were later collected using a vanadium foil (21.6~g in the beam).
%to normalize the detector response.

%A suite of detectors radially located at 4 m from the sample position record the TOF and position of each neutron event. The stored raw data, including normalization to beam intensity and instrument configuration, is then analyzed using the NIST package DAVE.

%The sample cell is a thin aluminum annulus with a 1.27 cm outer diameter and 0.9525 cm inner diameter, 11 cm tall for total volume 6.1 cc.  The annulus is formed by inserting a helium filled aluminum volume into the sample container. The beam was collimated to 9 cm of total height, and illuminated the entire width of the cell.

The data were reduced using the DCS MSLICE routine in the DAVE analysis suite developed by NCNR. Each data file was processed with proper background subtraction and detector efficiency normalization. We used the powder average technique, in which signals from any detectors with the same $2\theta$ diffraction angle were averaged and the final results were presented as a function of $|\vec{Q}|$ for each TOF bin (or $\omega$ bin). 
The processed data were re-binned and exported as tables of $S(Q,\omega)$ for further analysis.
%in PAN and Matlab. 
%The uncertainty associated with standard counting statistics.

\section{\label{sec:Elastic} Form Factors from Elastic Scattering}

\begin{figure}
\centering
(a)\includegraphics[width=3.5 in]{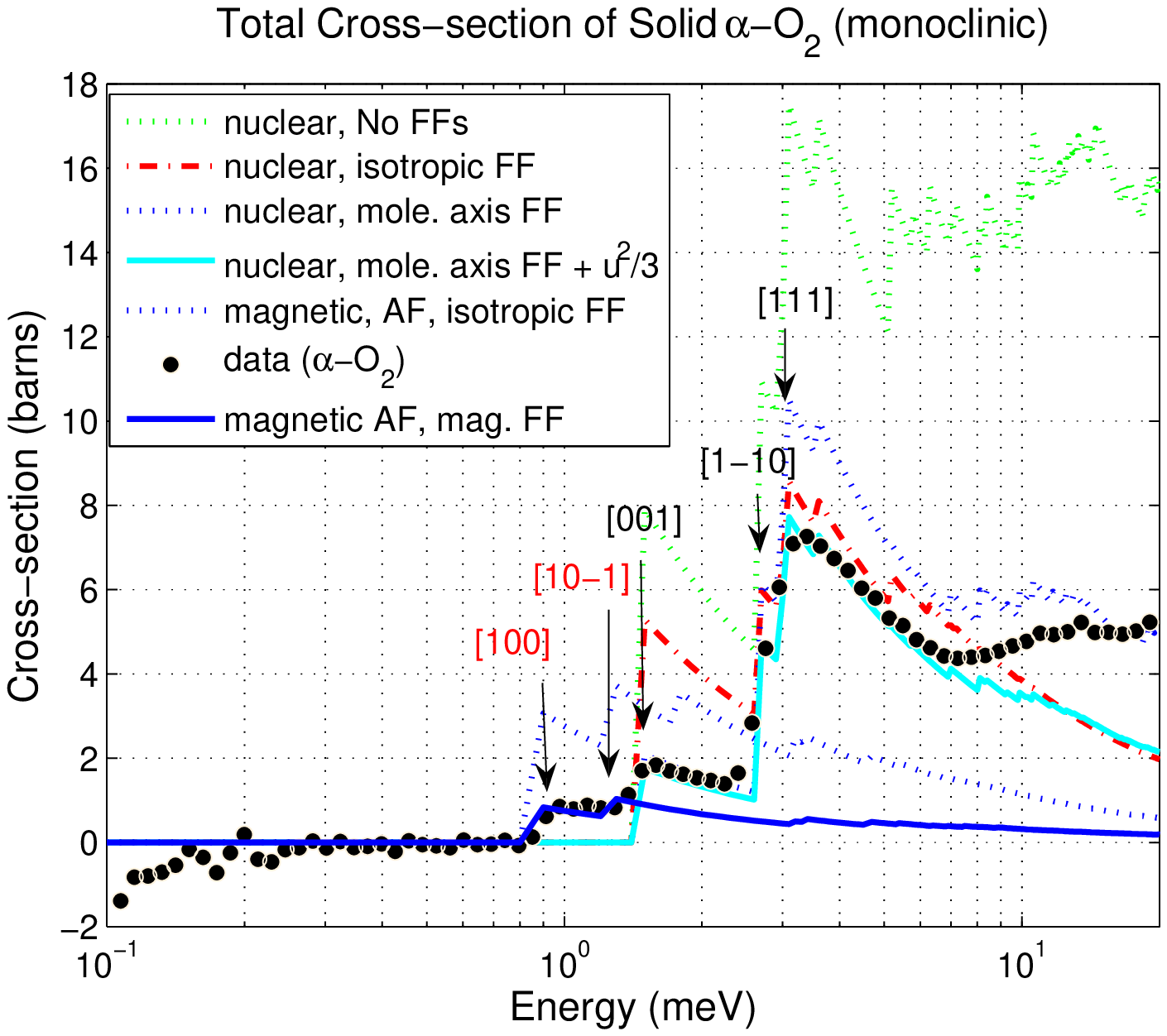}
(b)\includegraphics[width=3.45 in]{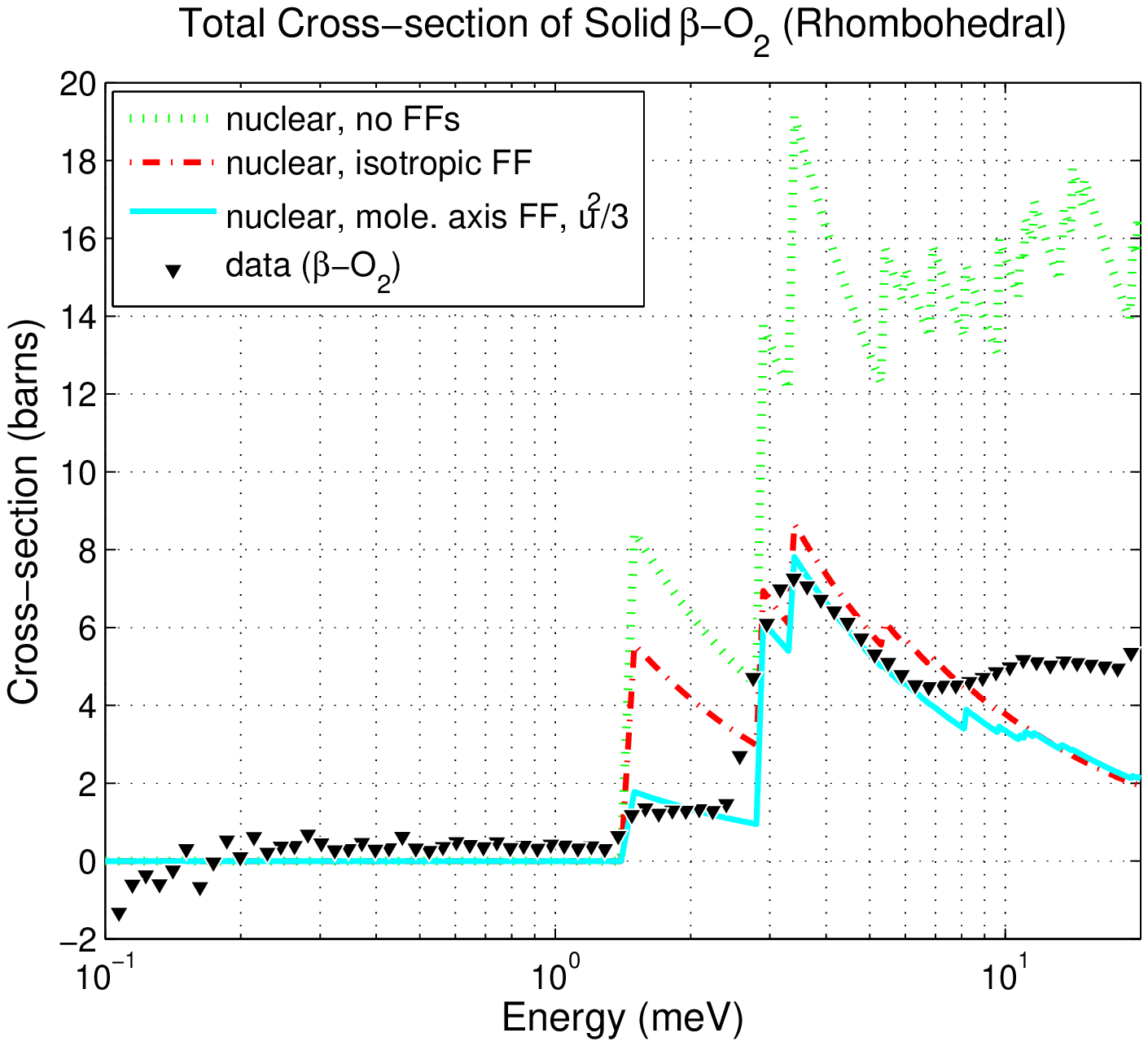}
\caption{\label{fig:TotalXS} Total cross-section of $\alpha$ and $\beta$-O$_2$.}
\end{figure}

\begin{figure}
\centering
\includegraphics[width=3.5 in]{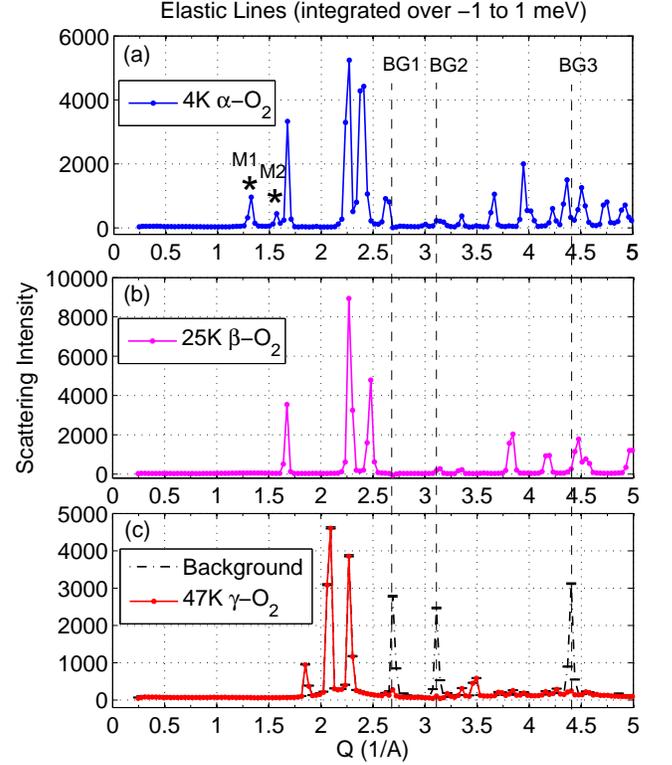}
\caption{\label{fig:Elastic} Neutron diffraction lines of s-O$_2$. The dashed lines are the 3 major background peaks from Al container cell are at Q=2.69, 3.11, and 4.37/$\mbox{\AA}$.}
\end{figure}

In order to get a correct handle of the distinct form factors in magnetic and nuclear scattering, required later to calculate the inelastic cross sections, we start by analyzing the data on elastic scattering. 
The total cross sections for $\alpha$- and $\beta$-O$_2$ are plotted in Fig.~\ref{fig:TotalXS} and the elastic diffraction lines are plotted in Fig.~\ref{fig:Elastic}. Note that the total cross section is predominantly elastic over the energy spectrum of the incident cold neutrons. 
The elastic cross section can be separated into coherent and incoherent contributions:
\begin{eqnarray}
\sigma^{coh}_0&=&4\pi\frac{\lambda^2}{8\pi V}\sum_{hkl}d_{hkl}|F^{coh}_{hkl}|^2e^{-2W}, \label{eqn:cohelastic} \\
\sigma^{inc}_0&=&4\pi |F^{inc}|^2 e^{-2W},
\end{eqnarray}
where $d_{hkl}$ is the inter-planar spacing between the [hkl] planes in the lattice space, i.e.,
\begin{equation}
d_{hkl}=\frac{2\pi}{|\vec G|}, \mbox{\hspace{0.2 in}where   } \vec G= h \vec g_1 + k \vec g_2 + l \vec g_3.
\end{equation}
Here $\vec{G}$ is the reciprocal lattice vector with integer numbers of $[h,l,k]$.
The form factor $F$ within the unit cell differs depending on whether the coherence of the scattered wave is retained:
\begin{eqnarray}
F^{coh}_{hkl} &=& \left\{ \begin{array}{ll}
\sum^{N_{basis}}_{i=1}b^{coh}_i e^{i \vec G \centerdot \vec R_i} & \mbox{if $\lambda \geqslant 2d_{hkl}$} \label{Eq:CohElas} \\
0 & \mbox{otherwise} \end{array} \right.   \\ 
F^{inc} &=&\sqrt{\sum^{N_{basis}}_{i=1} (b^{inc}_i)^2}.
\end{eqnarray}
In a non-Bravais crystal, the number of atoms in the unit cell, $N_{basis}$ can be more than 1.  

As shown in Fig.~\ref{fig:TotalXS}, form factors significantly modify the cross section in the following ways:
For nuclear scattering, the intra-molecular interference between the two atomic nuclei would introduce an additional molecular form factor,  
\begin{equation}
F^{mol.}=1+e^{i \vec{Q}\cdot \vec{r}},
\end{equation}
where $\vec{r}$ is the vector representing the intra-molecular separation of the two atomic nuclei. 
To facilitate the AF spin ordering, the O$_2$ molecular axis is aligned to be perpendicular to the basal plane.
%Because the molecular axis is preferentially pointing out of the basal plane for both $\alpha$ and $\beta$-O$_2$,
%aligned to be perpendicular to the basal plan, 
Hence, the vector $\vec{r}$ is represented as (0,0,$l$), with $l=1.21\mbox{\AA}$. Note that it would not be appropriate to apply the typical spherical Bessel function used in diatomic molecules having free rotations, such as H$_2$ and D$_2$~\cite{Liu2010}. Librational motions would lead to an additional Debye-Waller factor. The inclusion of $F^{mol.}$ reduces the elastic nuclear cross sections by about 50\%. 
%and is essential to reproduce the experimental data for both $\alpha$ and $\beta$ phases. 

For magnetic scattering, Stephens~\cite{Stephens85} and Dunstetter~\cite{Dunstetter96} suggested the magnetic form factor to be used for molecular O$_2$. In the $\alpha$ phase, we need to introduce a sign change in the unit-cell form factor in Eq.~\ref{Eq:CohElas} to reflect on the AF spin configuration. In a conventional unit cell with two O$_2$ molecules of opposite spin orientations, the lattice form factor becomes
\begin{equation} 
F^{coh}=b_0 e^{i \vec{Q}\cdot \vec{R}_1}-b_0 e^{i \vec{Q} \cdot \vec{R}_2},
\end{equation} 
where $\vec{R}_1=(0,0,0)$ and $\vec{R}_2$=(a/2,b/2,0) are the vector of the two basis in a monoclinic unit cell.  
The lattice constants are $a=5.375~\mbox{\AA}$, $b=3.425~\mbox{\AA}$, and $c=4.242~\mbox{\AA}$ for monoclinic $\alpha$-O$_2$, and reduce to  
$a=\sqrt{3.754^2+(3.425/2)^2}~\mbox{\AA}$ for rhombohedral $\beta$-O$_2$. 
As the AF long-range spin order is present only in the $\alpha$ phase, the change of sign does not apply to the $\beta$ phase. Doubling the periodicity of the spin lattice in $\alpha$-O$_2$ manifests as the additional Bragg edges at long neutron wavelengths in the total cross section (at the E$<$1.5~meV in Fig.\ref{fig:TotalXS}a) and magnetic diffraction peaks below 1.7~$\mbox{\AA}^{-1}$ (Fig.~\ref{fig:Elastic}a). 
The lowest magnetic Bragg cutoff at 0.8~meV results from the constructive interference between the [100] planes, which is the perpendicular to the direction of spin order. The second lowest magnetic Bragg edge corresponds to scattering along the [10$\bar{1}$] plane. Due to its small intensity and close proximity to the first nuclear Bragg edge, this second magnetic Bragg edge is harder to resolve in the total cross section, but can be clearly identified in the powder diffraction data. The two lowest $Q$ lines in the $\alpha$ phase (marked by the asterisks in Fig.~\ref{fig:Elastic}a, first reported by Collins~\cite{Collins66} as magnetic in origin) disappear in the $\beta$ phase, providing the evidence of long-range spin ordering in $\alpha$-O$_2$.  
Also note the similarity of the total cross section and the rest of the diffraction lines between $\alpha$ and $\beta$-O$_2$, indicating an adiabatic transformation of the nuclear lattice structures with some broken symmetries in the $\alpha$ phase.
We will take advantage of the similar lattice structures in $\alpha$ and $\beta$-O$_2$ to separate the magnetic excitations from the phonon excitations.
The $\gamma$-O$_2$ displays the symmetry of a cubic lattice (Fig.~\ref{fig:Elastic}c), with quite distinct diffraction patterns. 

\section{Dynamics from Inelastic Scattering}

\begin{figure}
\centering
\includegraphics[width=3.5 in]{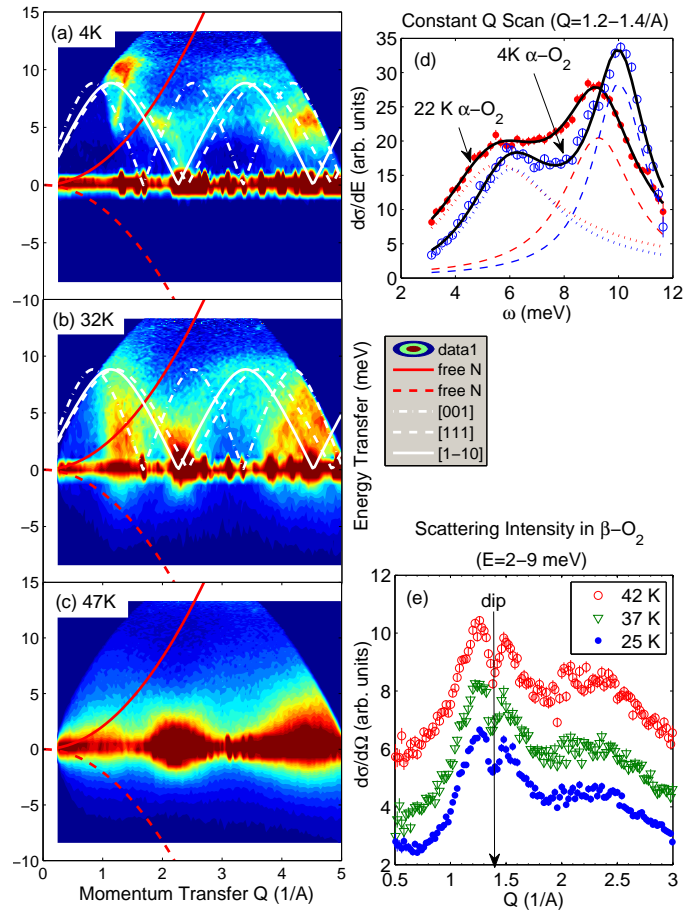}
\caption{The dynamic structure function $S(Q,\omega)$ of s-O$_2$ in the (a) $\alpha$ phase at 4~K, (b) $\beta$ phase at 32~K, and (c) $\gamma$ phase at 47~K, measured by INS. (d) Constant $Q$-slices of $S(Q,\omega)$ in $\alpha$-O$_2$ (e) $Q$-dependent amplitudes of integrated $S(Q,\omega)$ in $\beta$-O$_2$.}
\label{fig:SS_all}
\end{figure}

%\begin{figure}
%\centering
%%\includegraphics[width=3.5 in]{figures/Phonons_4K.eps}
%%\includegraphics[width=3.5 in]{figures/Phonons_32K.eps}
%%\includegraphics[width=3.5 in]{figures/GammaO2_SS.eps}
%\includegraphics[width=3.0 in]{figures/4K_SS_new.eps}
%\includegraphics[width=3.0 in]{figures/32K_SS_new.eps}
%\includegraphics[width=3.0 in]{figures/GammaO2_SS.eps}
%%\includegraphics[width=3.5 in]{figures/47K_SS_new.eps}
%\caption{\label{fig:DCSdata} Totoal Cross-section of $\beta$-O$_2$ oxygen}\end{figure}

The neutron scattering intensity is characterized by the double differential cross section, expressed by 
\begin{equation}
\frac{d\sigma}{d\Omega dE}=b_0^2\frac{k'}{k}S(\vec{Q},\omega).
\end{equation}
For polycrystalline samples, the dynamic structure function $S(\vec{Q},\omega)$ is measured as a powder average. 
In the limit that the orientations of micro-crystals are fully randomized inside the bulk sample, the angular dependence of the momentum transfer $\vec{Q}$ is averaged over 4$\pi$ to become $|\vec{Q}|\equiv Q$.
During each scattering, the incident neutron experiences a momentum transfer of $\vec{Q}=\vec{k}-\vec{k'}$, and an energy transfer of $\omega=E_i-E_f$. Processes with positive $\omega$ involve neutron downscattering with the neutron losing energy into creations of one or more quasi-particles in the s-O$_2$; processes with negative $\omega$ involves neutron upscattering with the neutron gaining an energy gain by annihilating one or more quasi-particles, absorbing all their energies. Two dimensional maps of $S(Q,\omega)$ at different temperatures were produced with on average 2 hours of neutron exposure. Samples of the $S(Q,\omega)$ maps are shown in Fig.~\ref{fig:SS_all} with color indicating the scattering intensity normalized by the phase space factor, $k/k'$.  
The INS measurements unfold the very distinct dynamics among the 3 solid phases.
%Because the energy of the scattered neutrons is determined by the time-of-flight of the neutron from the target cell to the detector, with finite timing resolution, the downscattering processes would have a higher energy resolution than the upscattering. 

%The raw data was processed using the DAVE program, and the converted into spe files. The spe files are analyzed using both PEN and matlab scripts.
%Neutron scattering probes fluctuations perpendicular to the scattering vector $\vec{Q}$.

\subsection{\label{sec:alpha} $\alpha$-O$_2$}

\begin{figure}
\centering
\includegraphics[width=3.5 in]{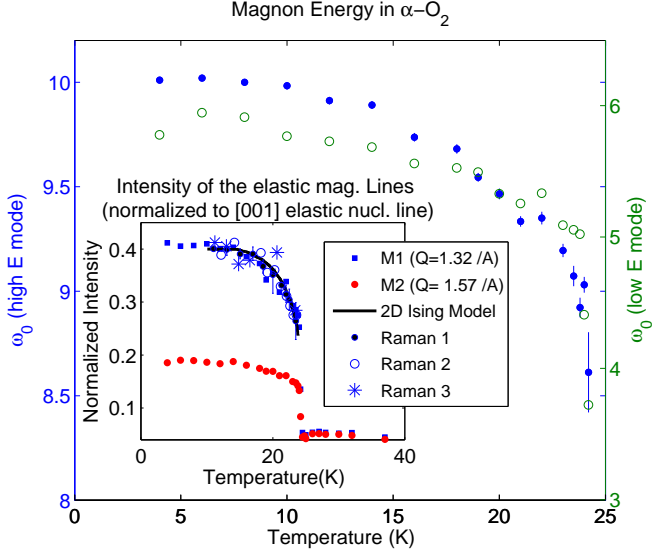}
\caption{\label{fig:MagElastic2} The central energy of the high energy and low energy magnetic clusters in $\alpha$-O$_2$. Inset: Intensity of the magnetic elastic lines in $\alpha$-O$_2$. }
\end{figure}

As we have discussed in sec.~\ref{sec:Elastic}, the long-range AF order is responsible for the two magnetic Bragg lines at $Q=1.32\mbox{ \AA}^{-1}$(M1) and $1.57\mbox{ \AA}^{-1}$(M2). To get a quantitative measure of the intensities of these elastic magnetic lines, we integrate the $S(Q,\omega)$ from $\omega=[-1,1]$~meV, and normalize to that of the first nuclear Bragg peak at $Q=1.7\mbox{\AA}^{-1}$. The temperature dependence of the intensity of M1 can be compared with previous optical measurements~\cite{Stephens86} (Fig.~\ref{fig:MagElastic2} inset), and shows good agreements with the prediction of 2D Ising Model. The decreasing intensity with increasing temperature is consistent with the thermally induced reduction of the spin-spin correlation predicted in the Monte-Carlo simulation~\cite{LeSar88}.  

%To facilitate the AF spin ordering, the O$_2$ molecules axis are aligned to be perpendicular to the basal plane. 
%The molecular libration is negligible in the $\alpha$ phase, probably because of the strong spin coupling.
Following the work by Stephens {\it et al.}~\cite{Stephens86}, we describe the
magnetic excitations in $\alpha$-O$_2$ with a Heisenberg spin Hamiltonian:
\begin{equation}
H=-2\sum_{<ij>}J_{ij}\vec{S}_i\cdot \vec{S}_j - D \sum_{i} S_{xi}^2-D'\sum_i (S_{yi}^2-S_{zi}^2),
\end{equation} 
based on the renormalized spin wave theory previously developed for NiCl$_2$~\cite{Lindgard75}, which is a simple AF system with a rhombohedral structure. 
The exchange interactions in solid $\alpha$-O$_2$ include the 4 nearest neighbors, $J_{NN}$, and 2 next nearest neighbors, $J_{NNN}$, all within the basal plane. The inter planar coupling is negligible due to the large planar separation.
%Monte-Carlo simulations~\cite{LeSar1988} also confirmed this spin correlations between the nearest neighbors. 
The $x$ axis is the easy magnetization direction (along the crystal $b$ axis), along which the spins align co-linearly. 
The $y$ axis points along the crystal $a$ axis, and perpendicular to the basal plane is the $z$ axis, along which aligns the molecular axis. 
The Hamiltonian contains the single particle term with a self-energy $D$, introduced to characterize the spin anisotropy.
The Ising-like behavior, as revealed by the temperature dependent intensity of the magnetic Bragg peak, strongly suggests that that the anisotropies shall satisfy $D>0$ and $D'\simeq 0$. If $D=D'>0$, we would expect the system to have an easy plane magnetization. 

\begin{figure}
\centering
\includegraphics[width=3.3 in]{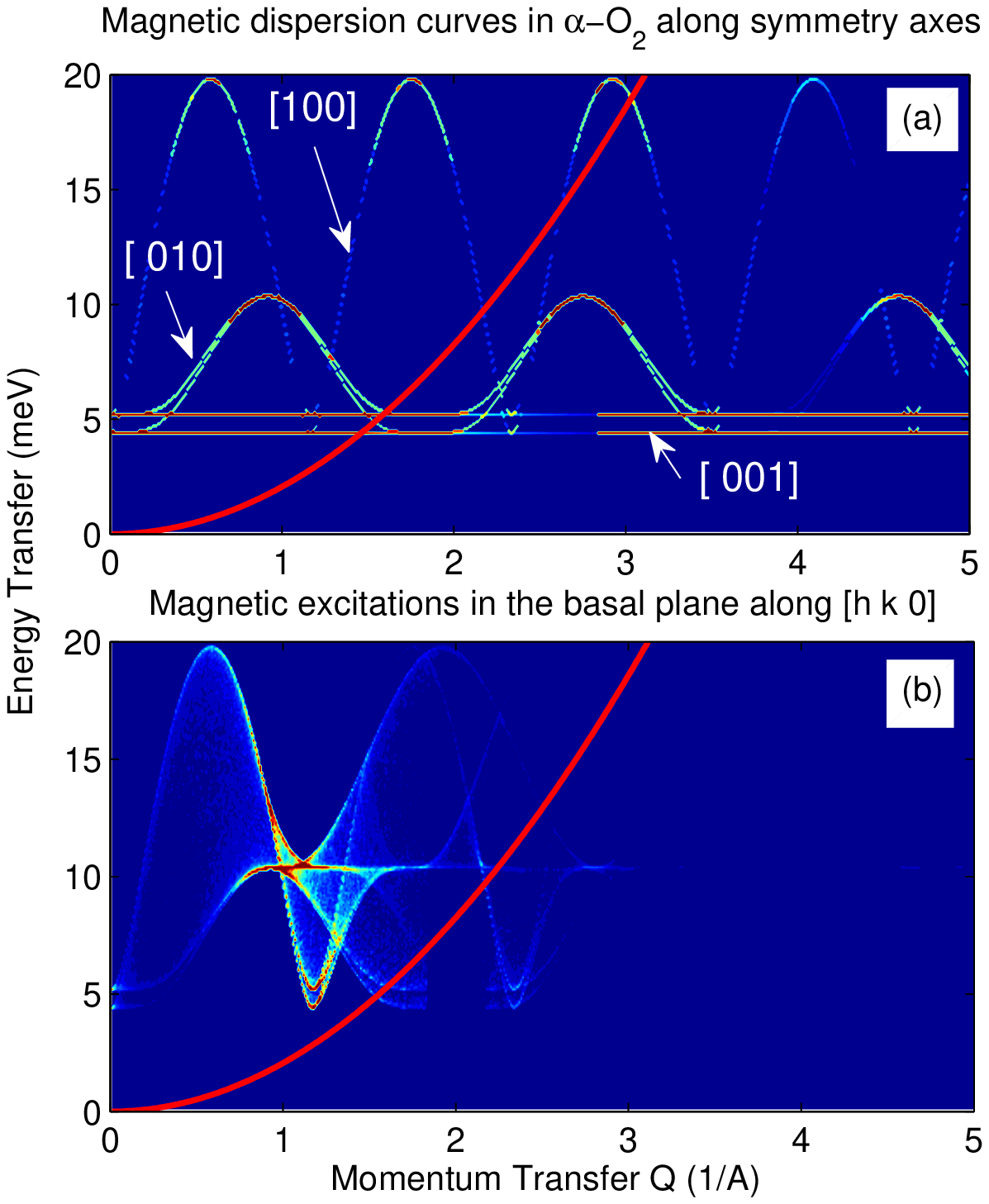}
\includegraphics[width=3.3 in]{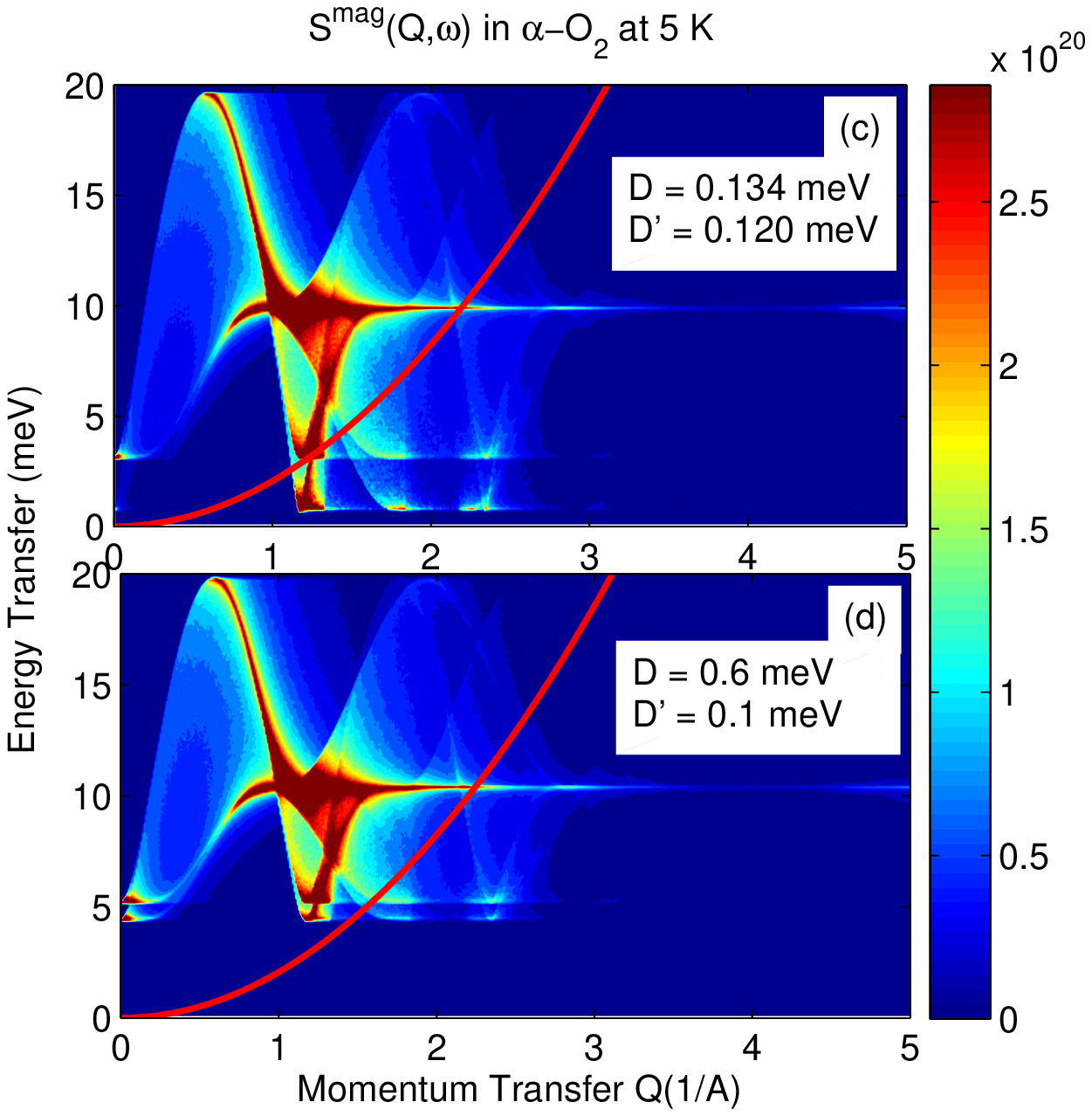}
\caption{\label{fig:dispersion} Calculated $S(Q,\omega)$ of coherent magnetic scattering in $\alpha$-O$_2$ using the magnon energy dispersion relation derived from the normalized spin wave theory. (a) along symmetry axes (b) inside the basal plane (c) polycrystalline average with small anisotropy energies (d) polycrystalline average with a large anisotropy energy.}
\end{figure}

Using the renormalized spin wave theory, the magnon energy spectrum of this two-dimensional $S=1$ system is solved~\cite{Stephens86,Lindgard75}:
\begin{align}
&E^2_a=(2J_{NNN}(0)-2J_{NNN}({\vec{q}})-2J_{NN}(0)+D)^2 \nonumber \\
& \hspace{1. in} - [2J_{NN}({\vec{q}})+(-1)^aD']^2,
\label{Eq:dispersion}
\end{align}
with the index $a=0$ as the acoustic mode and $a=1$ the optic mode. The coupling $J_{NN}({\vec{q}})=\sum_i J_{NN} e^{i\vec{q}\cdot\vec{r}_i}$ is summed over the nearest neighbors on the opposite sublattice, and $J_{NNN}$ is the next nearest neighbors on the same sublattice. 
The best value of the coupling constants reported by Stephens~\cite{Stephens86} are: $J_{NN}=-2.44$~meV, $J_{NNN}=-1.22$~meV, $J_{\perp}=0$, $D=0.134$~meV, and $D'=0.120$~meV, which suggests an easy-plane magnetization (a XY-model). As we will show next, the coupling constants should be modified, in order to bring the model in better agreements with the new high resolution INS data and to conform with the Ising model suggested by the elastic scattering data. 
Fig.~\ref{fig:dispersion}a illustrates
the energy dispersion relations calculated using the updated coupling constants along major symmetry axes. Not surprisingly, there exists a finite energy gap for magnon excitations even at $Q=0$, as a result of the anisotropy.
As we allow the scattering to be confined in the basal plane, the $S(Q,\omega)$ starts to cluster around regions with modes crossing (Fig.~\ref{fig:dispersion}b) along the [100] direction. 

As neutron scattering probes fluctuations perpendicular to the scattering vector $\vec{Q}$, the magnetic dynamic structure function of a AF system can be written as 
\begin{align}
&S^{mag}(\vec{Q},\omega)=(1+\hat{Q}_b^2)[u_q+(-1)^mv_q]^2\delta(E_a(\vec{q})-\hbar\omega) \nonumber \\
&\hspace{1.5 in} \times \delta(\vec{Q}-\vec{q}-\vec{G}).
\end{align}
Here the scattering amplitude is enhanced when the momentum transfer points along the spin aligned axis, i.e. the $b$ axis.
The delta functions enforce the energy and momentum conservation in this coherent scattering process.
The Holstein-Primakoff transformation coefficient used in Stephens' paper~\cite{Stephens86}, when 
compared against the experimental data of $S(Q,\omega)$, is found  
to carry an incorrect sign, and should be corrected as the following:
\begin{align}
&[u_q+(-1)^mv_q]^2 \nonumber \\
&=2S\frac{(2J_0-2J'_0-2J_q+D)+(-1)^m[2J'_q+(-1)^aD']}{E_q(\vec{q})}.
\end{align}
It was this sign error together with the low values of anisotropy energies that led to 
%the non-physical energy dispersion with certain range of coupling constants. It also pushed the 
predictions of magnon modes with energies lower than the result of the new INS measurements. This led us to over-predict the UCN yields when we first proposed the idea of magnetic UCN production using $\alpha$-O$_2$~\cite{Liu2004}.   

With the $Q$-dependent energy dispersion relation in Eq.~\ref{Eq:dispersion}, we calculate the dynamics structure function for magnon excitations in polycrystalline $\alpha$-O$_2$, following the algorithm we have developed to calculate coherent inelastic scattering cross section in solid deuterium~\cite{Liu2010}. The magnetic form factors on the molecular and lattice level have already been validated using the elastic cross section as discussed in Sec.~\ref{sec:Elastic}. With the coupling constants suggested by Stephens {\it et.al.}~\cite{Stephens86}, the calculated $S^{mag}(Q,\omega)$ is shown as an intensity plot in Fig.~\ref{fig:dispersion}c. 
Because of the molecular spin form factor, the scattering amplitude becomes negligible for $Q>2\mbox{\AA}^{-1}$, above which the phonon scattering dominates. The small anisotropy energies lead to a soft spectrum (clustered around the center of the second Brillouin zone) extending down to energies as low as $\sim 0.8$~meV. 
If $D=0$, the excitation would become gapless at $Q=1.3\mbox{\AA}^{-1}$. If $D=D'>0$, the excitation also becomes gapless. It requires the adjustment of the easy axis anisotropy energy up to $D=0.6$~meV, to bring the low energy mode higher than 5~meV to better agree with the experimental data (Fig.~\ref{fig:SS_all}a). This again supports the conclusion that the nature of the magnetism in $\alpha$-O$_2$ is indeed 2D Ising. 
The increased anisotropy energy also agrees with the free gas value of 5.712~K, that characterizes the inter molecular spin-orbit and spin-spin coupling. The single spin anisotropy term is A=3.96 cm$^{-1}$. 
When considering the molecular librations, the spin-libration interaction can only soften the anisotropy by 10\%~\cite{Freiman04}. 

Varying the value of $J_{NN}$, on the other hand, shifts the energy of the 19~meV magnons (corresponding to excitations at the Brillouin zone boundary), which is beyond the range of energy coverage of the measurement.  The difference $J_{NN}-J_{NNN}$ determines the placement of the dominant magnon at 10~meV at the zone boundary. Results of our numerical studies agree with Stephens' suggestion of $J_{NN}-J_{NNN}=-1.22$~meV.

\begin{figure}
\centering
\includegraphics[width=3.3 in]{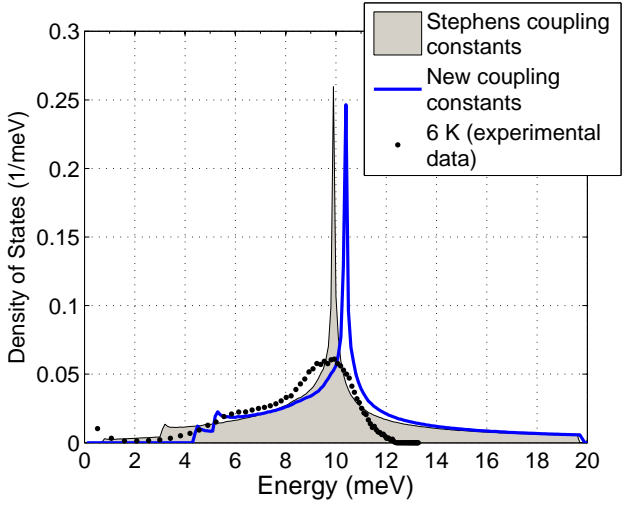}
\caption{\label{fig:ZZ_alpha} The density of states of magnons in $\alpha$-O$_2$.}
\end{figure}

We calculate the density of states of the magnetic excitations by inverting the dynamic structure function as detailed in later discussions (sec.~\ref{sec:ZZ}). As shown in Fig.~\ref{fig:ZZ_alpha}, increasing the easy-axis anisotropy energy significantly shifts the population of low energy magnon modes towards a higher energy gap, whereas the 10~meV mode shifts only by 5\%. 
The density of states derived from experimental data (limited to magnetic excitations at $Q<1.7\mbox{\AA}^{-1}$) is plotted for comparison. Because of the limited range of coverage in the energy transfer at low $Q$, the data cuts off around 12~meV. Except for the sharp peak around 10~meV (which is washed out by the finite energy resolution no better than 0.5~meV), the measured density of state of magnetic excitations is in general agreement with the results of the 2D Ising AF spin-wave model, with a large anisotropy energy.

For $Q=1.2\sim 1.4\mbox{\AA}^{-1}$, the scattering is predominantly magnetic. We fit $S(Q,\omega)$ over the constant Q slice as a sum of a Lorentzian (centered around 10~meV) peak and a phenomenological 3-pole response function Eq.~\ref{eq:diffuse} (as shown in Fig.~\ref{fig:SS_all}d).
As the solid temperature increases, both the high energy cluster (E$\sim$10~meV) and the low energy cluster (E$\sim$5~meV) shift downwards in energy (see Fig.~\ref{fig:MagElastic2}, in accord with the thermally induced reduction in exchange couplings, as previously shown with the intensity of the elastic magnetic lines (Fig.~\ref{fig:MagElastic2} inset). 

%\begin{figure}
%\centering
%%\includegraphics[width=3.6 in]{figures/Scaling_alpha.eps}
%\includegraphics[width=3.5 in]{figures/Magnon_energy_alpha.eps}
%\caption{\label{fig:Qscan} Scattering intensity at constant $\omega$ in O$_2$ oxygen}
%\end{figure}

%\begin{figure}
%\centering
%\includegraphics[width=3.3 in]{figures/SSO2_mag_compare.eps}
%\caption{\label{fig:mag_alpha} Magnetic cross-section of $\alpha$-O$_2$ oxygen}\end{figure}

%\begin{figure}
%\centering
%\includegraphics[width=3.2 in]{figures/Magnon_energy_alpha.eps}
%\includegraphics[width=3.5 in]{figures/Magnon_energy_alpha_combine.eps}
%\caption{\label{fig:MagElastic} Intensity of the elastic line in the $\alpha$-O$_2$ oxygen}
%\end{figure}

%\begin{figure}
%\centering
%\includegraphics[width=3.3 in]{figures/SSO2_mag_fine2.eps}
%\caption{\label{fig:TotalXS_alpha} Totoal Cross-section of $\beta$-O$_2$ oxygen}\end{figure}

The Neel temperature T$_N=\frac{2}{3}S(S+1)J(0)=\frac{4}{3}(4J_{NN}-2J_{NNN})$, calculated within the mean field approximation, is 35~K. This is about 1.5 times higher than the $\alpha-\beta$ transition temperature of 24~K. Inclusions of the molecular field, however, could not explain the discrepancy. Perhaps the quantum fluctuations prevent the spin from ordering until the lower temperature is reached.
The exchange field can be estimated using 
\begin{equation}
B_{exch.}=\frac{2zJ_{NN}}{g\mu_BS}=227 \mbox{ T}
\end{equation}
is quite large. With the measured spin flop field of 7~T~\cite{Uyeda85} 
\begin{equation}
B_{sf}^2=2B_{anis.}B_{exch.},
\end{equation}
we arrive at the anisotropy field of about 0.11~T. 

\subsection{$\beta$-O$_2$}

As revealed with the INS measurements (Fig.~\ref{fig:SS_all}), $\beta$-O$_2$ has a magnetic excitation that differs significantly from the magnons in the $\alpha$ phase. First, the intensity of the magnetic Bragg line at $Q=1.3\mbox{ \AA}^{-1}$ drops by orders of magnitude when compared to that in the $\alpha$ phase.
However, there appears a prominent magnetic excitation centering broadly around $Q=1.3\mbox{ \AA}^{-1}$. The excitation is dispersion-less in energy and extends over $\omega=0 \sim 8$~meV. 
%However, the intensity of the magnetic Bragg line at $Q=1.3\mbox{ \AA}^{-1}$ drops by orders of magnitude when compared to that in the $\alpha$ phase.  
%When integrating the scattering intensity from 2 to 9~meV, the $d\sigma/d\Omega$ shows a sharp dip around the center of the peak. This was not reported in any of the previous studies. This is the artifact of dead detectors. 
%This can be compared to the diffuse scattering in 
%the liquid and soild $\gamma$ phase of oxygen~\cite{Fernandez08}. 
When compared to the spin diffusion dynamics in $\gamma$-O$_2$ at low $Q<1.5\mbox{\AA}^{-1}$~\cite{Fernandez08}, this dispersion-less magnetic excitation in $\beta$-O$_2$ looks intrinsically different as it does not extend continuously to low $Q$s as in $\gamma$-O$_2$. 
%the peak around $Q\sim 1 \mbox{ \AA}^{-1}$, half the momentum transfer where the main peak of the magnetic structure factor found in both phases, is attributed to the magnetic scattering~\cite{Fernandez2008}. 
%In comparing INS data below 1.5$\mbox{\AA}^{-1}$ between the liquid phase and $\gamma$-phase, they conclude that the diffuse features have the same magnetic origin. 
%By studying the second moment, the authors derived the best value for the coupling constants $J_{NN}=0.84\pm 0.03$ meV, and $J_{NNN}=0.11 \pm 0.07$ meV in $\gamma$-O$_2$. The exchange constants derived in the liquid is $J_{NN}=0.41 \pm 0.05$ meV, about half the strength of that in the solid $\gamma$ phase. 
%The data set below $Q=0.7 \mbox{\AA}^{-1}$ in both phases indicates a dynamic of spin diffusion with a distance-time scale of 0.14 $\AA ps^{-1}$, which is one order of magnitude faster than the characteristic translational and rotational motion measured in the liquid. The reorientational correlation time scale is about two orders of magnitude slower than these values. 
%The instability $Q$ at 1.32 $\mbox{\AA}$ corresponds to a length scale of 4.76 $\mbox{\AA}$. This is roughly $\sqrt{2}$ of the lattice spacing between the nearest neighbors on a triangular lattice. 
There are many attempts to derive the spin configuration and its short-range order by studying the detailed structure of the broad elastic peak, based on different variations of the three-sublattice Yaffet-Kittel structure~\cite{Rastelli86, Rastelli88}. We would like to point out that 
the spin-pairing to form a spin singlet could also account for the vanishingly small elastic scattering amplitude. However, high quality data using single crystals are required to definitely support the spinon formation in $\beta$-O$_2$.

\begin{figure}
%\centering
\includegraphics[width=3.5 in]{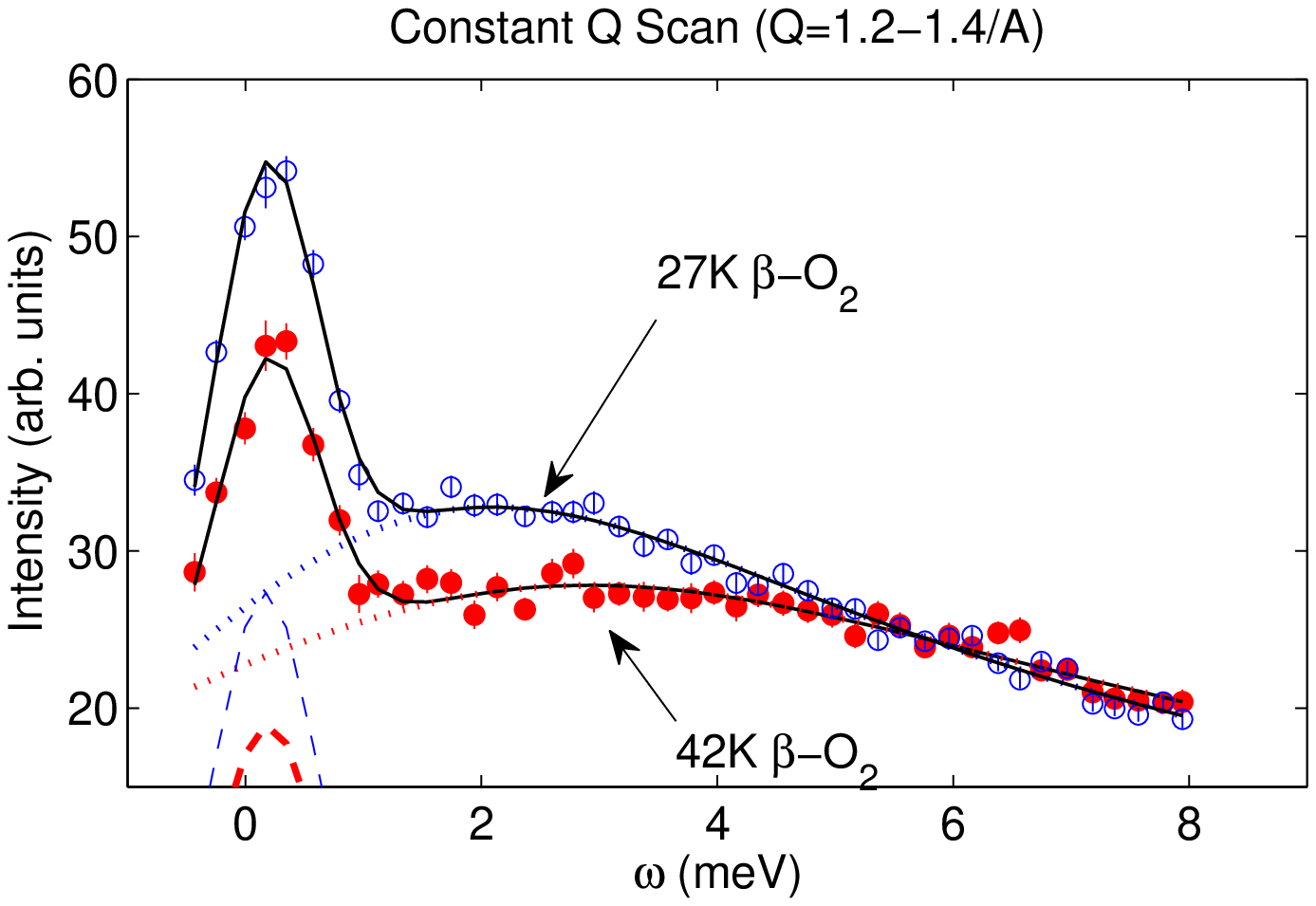}
\includegraphics[width=3.5 in]{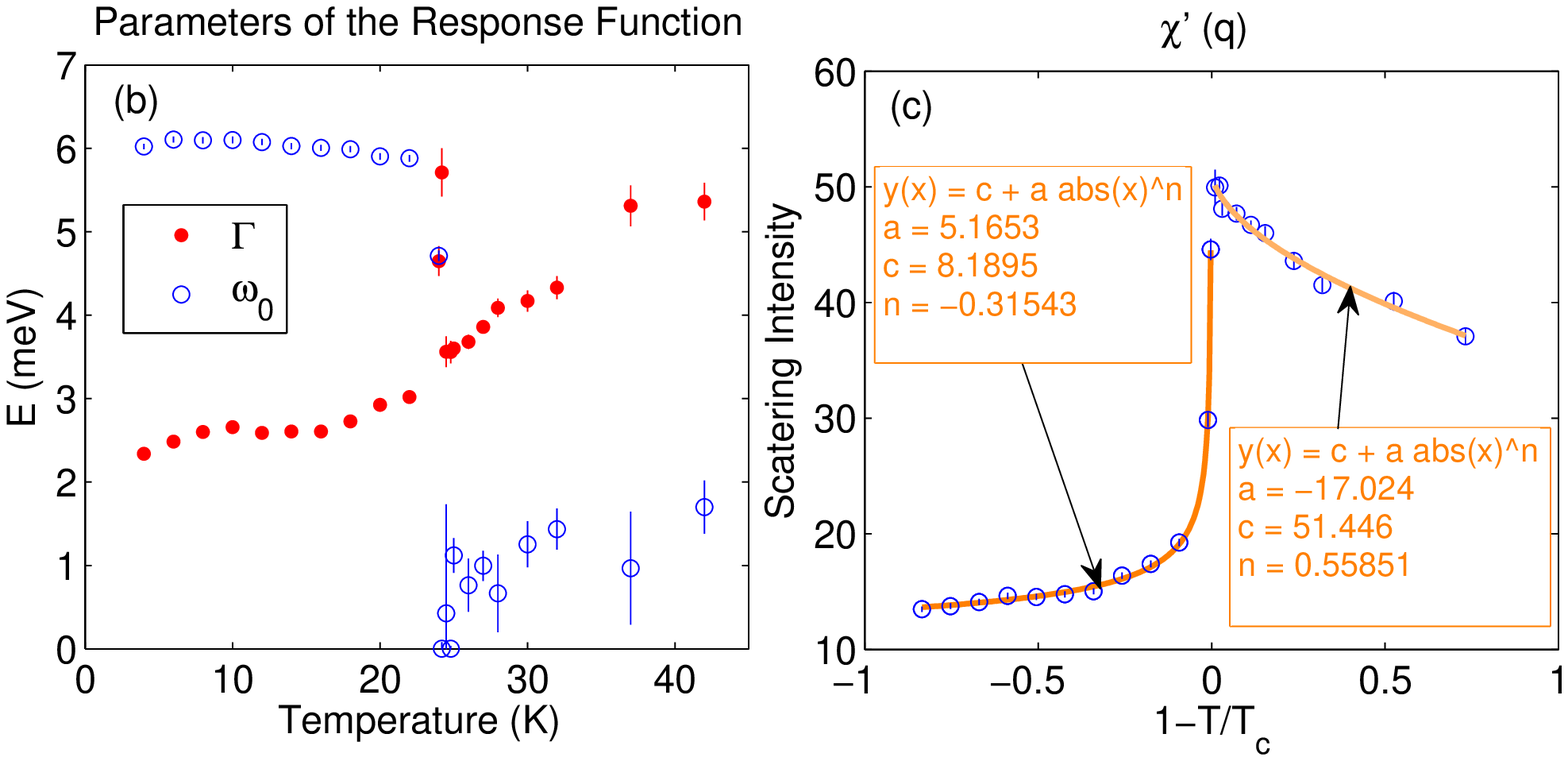}
\caption{\label{fig:Mag_beta} Inelastic magnetic scattering in $\beta$-O$_2$ fitted with the phenomenological 3-pole response function (a). (b) The temperature dependence of the fitted parameters. (c) The temperature scaling of $\chi'(q)$.}
\end{figure}

Using the dissipation-fluctuation theorem, the amplitude of inelastic scattering in a magnetic system can be described as the imaginary part of the magnetic susceptibility $\chi(q)=\chi'(q)+i\chi''(q)$. 
To characterize the energy dependence of this scattering, we fit the scattering amplitude (summed over $Q=1.2\sim 1.4 \mbox{\AA}^{-1}$) using the phenomenological 3-pole response function,
\begin{align}
&\chi''(q,\omega)= \frac{\chi'(q)}{2}\omega \times \nonumber \\
& \left(\frac{\Gamma(q)}{(\omega-\omega_0(q))^2+\Gamma^2(q)} + \frac{\Gamma(q)}{(\omega+\omega_0(q))^2+\Gamma^2(q)} \right),
\label{eq:diffuse}
\end{align}
where $\chi'(q)$ is the $q$-dependent susceptibility, $\Gamma(q)$ is the inter-site fluctuation rate, and $\omega_0(q)$ is a phenomenological energy pole. Examples of the fit and the temperature dependence of the fit parameters are shown in Fig.~\ref{fig:Mag_beta}.

Most of the early literatures used the ``coupled paramagnet" model to explain the large Lorentzian type of dynamic response and the $q$-dependence.
However, the rhombohedral structure in $\beta$-O$_2$ is equipped with all necessary conditions of a geometrically frustrated AF magnet.
We attribute the observed dispersion-less energy excitations to the unique spin dynamics in geometrically frustrated spins. 
The Curie-Weiss temperature is between 61~K to 55~K (using the coupling constants measured in the $\gamma$ phase), larger than the Neel temperature T$_N$ of 35~K.
The frustration parameter, defined as $|\Theta_{CW}|/\mbox{T}_N = 2.3\sim2.5$, is large. 
In temperatures between T$_N$ and $|\Theta_{CW}|$, the quantum fluctuation of the spin could lead to the spin liquid state. The formation of the quantum spin liquid requires dominant direct exchange, that can be described by Heisenberg Hamiltonian. The intrinsic fluctuations lead to a small coherence scale. In addition to quantum fluctuations, the spin-orbit coupling and inter-plane coupling can both give rise to fluctuations, lifting the degeneracy of the ground state.
The bandwidth of the energy excitation is on the order of $|\Theta_{CW}|$ = 4.7 $\sim$ 5.3 meV, which agrees with the fitted values of $\Gamma$ (Fig.~\ref{fig:Mag_beta}b) and is too big to be accounted for by thermal fluctuations alone.  
As an exercise, we could attempt to extract the scaling law of this transition by fitting $\chi'$ with a power law along both sides of the transition temperature (Fig.~\ref{fig:Mag_beta}c). The critical exponent is about 1/2 when approaching the transition from above, and about -1/3 when approaching the transition from below.

The $Q$ dependence of the elastic scattering intensity (Fig.~\ref{fig:SS_all}e) is best described by a short range, two dimensional helicoidal order with a correlation length of 5 $\mbox{\AA}$, corresponding to the next nearest neighbor on the {\it ab} basal plane. The three-sublattice Yaffet-Kittel model is thus ruled out. The dip on the inelastic scattering intensity shown in Fig.~\ref{fig:SS_all}e (summed over 2$\sim$9~meV) is an artifact of a dead detector, otherwise, it shows the broad peak centering around $Q=1.3\mbox{\AA}^{-1}$. 
The $Q$ width of the elastic magnetic peak decreases by about 20\% as the $\beta$-O$_2$ cools from 44~K to 24~K. 
%The $Q$ dependence of the inelastic scattering intensity (integrated over different ranges of $\omega$) is shown in Fig.~\ref{fig:Eint_beta}.

%1. The elastic line has a weak amplitude, but the correlation length can still be extracted..... How does this correlation length change with temperature?

%2. The energy width of the quisi-elastic is very large $\tau$~5 meV >$k_BT$.
%Usually width is ....

%Most of the literature claim that this is ``coupled paramagnet" to explain the large Lorentzian type of dynamic response, and the $q$-dependence.
%However, the rhomboheral structure is a perfect realization of the geometrically frustrated AF magnets, mostly observed at low temperatures. 

Due to the complexity of the spin dynamics in $\beta$-O$_2$,
we don't yet have an energy dispersion relation with which to construct a complete map of $S(Q,\omega)$, like what we have done for $\alpha$-O$_2$. 
%The direct comparison awaits for better understanding of th.

%What is the scaling law of the magnetic phase transition. What is fluctuation in the frustrated magnet? Static ($q$ dependence) or dynamic ($\omega$ dependence)?
%7. The fluctuation observed in the gamma phase is due to rotational diffusion, is there any magnetic component to the dynamic response?
%8. The gamma-beta phase transition is from orientational disorder to order transition with a large change in the molar volume! Magnetoelastic transition?

%\begin{figure}
%\centering
%\includegraphics[width=3.6 in]{figures/QScan.eps}
%\caption{\label{fig:Qscan} Scattering intensity at constant $\omega$ in O$_2$ oxygen}\end{figure}

%\begin{figure}
%\centering
%\includegraphics[width=3.0 in]{figures/Eintegrate_beta.eps}
%\caption{\label{fig:Eint_beta} Totoal Cross-section of $\beta$-O$_2$ oxygen}
%\end{figure}

%\begin{figure}
%\centering
%%\includegraphics[width=3.6 in]{figures/lowQ_sus.eps}
%%\includegraphics[width=3.6 in]{figures/lowQ_width.eps}
%%\includegraphics[width=3.2 in]{figures/Scaling_beta.eps}
%\includegraphics[width=3.5 in]{figures/ResponseFunction_beta.eps}
%\caption{\label{fig:Qscan} Scattering intensity at constant $\omega$ in O$_2$ oxygen}
%\end{figure}

%\begin{figure}
%\centering
%\includegraphics[width=3.0 in]{figures/Eintegrate.eps}
%\caption{\label{fig:Eint_beta} Totoal Cross-section of $\beta$-O$_2$ oxygen}
%\end{figure}

\section{\label{sec:ZZ} Evolution of the dynamics}

\begin{figure}
\centering
%(a).\includegraphics[width=3.0 in]{figures/T_dep_Q_mag1.eps}
%(b).\includegraphics[width=3.0 in]{figures/T_dep_Q.eps}
\includegraphics[width=3.3 in]{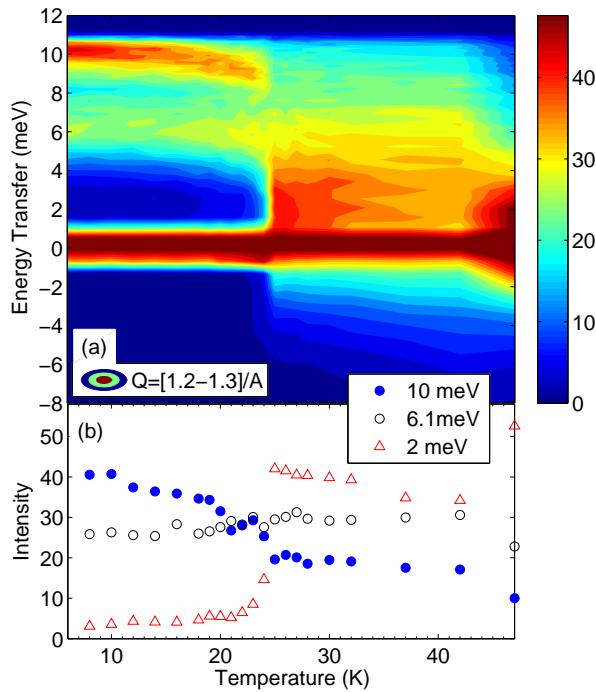}
\caption{\label{fig:Tdep_mag} The temperature evolution of the magnetic dominant $S(Q,\omega)$ (integrated over $Q=1.2\sim 1.3 \mbox{\AA}^{-1}$).}
\end{figure}

%\begin{figure}
%\centering
%%(a)\includegraphics[width=3.5 in]{figures/Escan_T_mag.eps}
%%(b)\includegraphics[width=3.5 in]{figures/Escan_T_nuc.eps}
%\includegraphics[width=3.3 in]{figures/Escan_T_both.eps}
%\caption{\label{fig:Escan_mag} Scattering intensity at constant $Q$.}
%\end{figure}

%\begin{figure}
%\centering
%\includegraphics[width=3.3 in]{figures/T_dep_ph.eps}
%\caption{\label{fig:Tdep_ph} Nuclear scattering intensity integrated over $Q=1.9\sim 2.0 \mbox{\AA}^{-1}$}
%\end{figure}

%\begin{figure}
%\centering
%(a)\includegraphics[width=3.3 in]{figures/MagTempDep_2.eps}
%(b)\includegraphics[width=3.3 in]{figures/ExcitationIntensity.eps}
%\caption{\label{fig:ZZ} Density of states in solid O$_2$}
%\end{figure}

Up to now, we have limited our discussions on the magnetic excitations, which are dominating the dynamics for $Q$ below 1.5$\mbox{\AA}^{-1}$ within the kinematic range covered by our INS measurements. Even without the polarization analysis, we can safely isolate the magnetic components by taking advantage of the similar lattice structure and lattice spacings between $\alpha$ and $\beta$-O$_2$, which warrant the overall similarity of the translational excitations (phonon) between these two phases. 
To demonstrate this, we plot the dispersion curves of acoustic phonon branches along major symmetry axes on the $S(Q,\omega)$ maps of $\alpha$ and $\beta$-O$_2$ (Fig.~\ref{fig:SS_all}a, b). These phone dispersion curves clearly outline the regions where the phonon excitations dominate. In particular, within the second Brillouin zone, one can clearly see the typical excitations of acoustic phonons extending all the way down to zero $\omega$ at $Q=2.2\sim 2.4 \mbox{\AA}^{-1}$, connecting to the elastic nuclear Bragg lines (Fig.~\ref{fig:Elastic}a, b).  
The phonon excitations within the first Brillouin zone are out of the assessable kinematic range. At larger $Q$s, the multiplicities of the inverse lattice planes increases, and it becomes harder to identify the specific phonon branches as they overlap. 

\begin{figure}
\centering
\includegraphics[width=3.5 in]{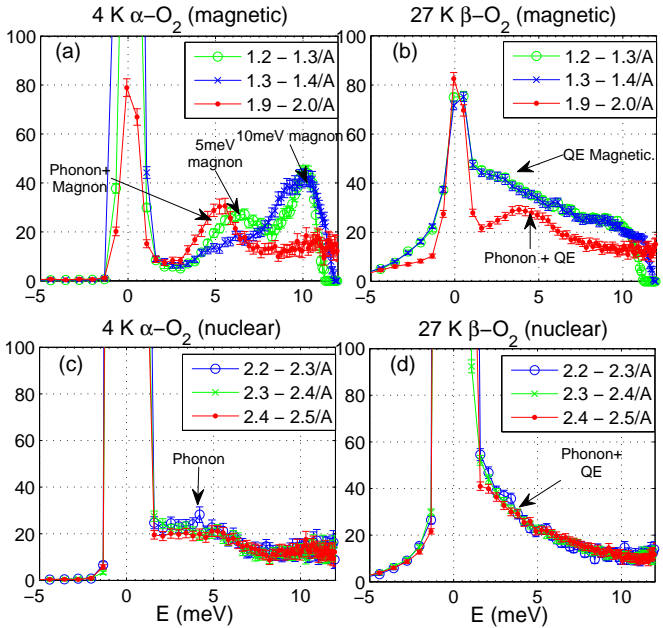}
\caption{\label{fig:Escan_mag} Constant $Q$ slices of $S(Q,\omega$ in $\alpha$-O$_2$ at 4~K and $\beta$-O$_2$ at 27~K. Magnetic excitations (a, b) and phonon excitations through nuclear scattering (c, d). } 
\end{figure}

Furthermore, the magnon and phonon dynamics can be well separated because the molecular spin form factor suppresses the scattering intensity at large $Q$, limiting the magnetic excitations to $Q<2\mbox{\AA}^{-1}$. To study the evolution of spin dynamics over solid phases, we integrate $S(Q,\omega)$ over $Q=1.2\sim 1.3\mbox{\AA}^{-1}$, and plot the scattering intensity as a function of $\omega$ and temperature (Fig.~\ref{fig:Tdep_mag}).
As the solid cools through the $\beta$-$\alpha$ transition, the magnetic excitation transforms from the broad-band dispersion-less excitations into the well defined magnon excitations consisting of a high energy mode around 10meV and a low energy mode around 6 meV, leaving almost no scattering amplitudes at low $\omega < 4$~meV in the $\alpha$ phase. The same details are illustrated in Fig.~\ref{fig:Escan_mag}a, b comparing the scattering amplitudes between $\alpha$-O$_2$ at 4~K and $\beta$-O$_2$ at 27~K. 

The evolution of the phonon dynamics can be examined by plotting $S(Q,\omega)$ at constant $Q$ slices at $Q=1.9 \sim 2.5 \mbox{\AA}^{-1}$ (Fig.~\ref{fig:Escan_mag}c, d). Solid $\beta$-O$_2$ shows more quasi-elastic scattering around the phonon branches when compared to the $\alpha$ phase. This might be a result of the additional spin-orbit and inter-plane couplings in the $\beta$ phase transferring the high degree of spin fluctuations to the lattice vibration.    
Beyond the additional quasi-elastic scattering amplitude, there are not much noticeable differences in the phonon dynamics between these two phases. 

\begin{figure}
\centering
\includegraphics[width=3.3 in]{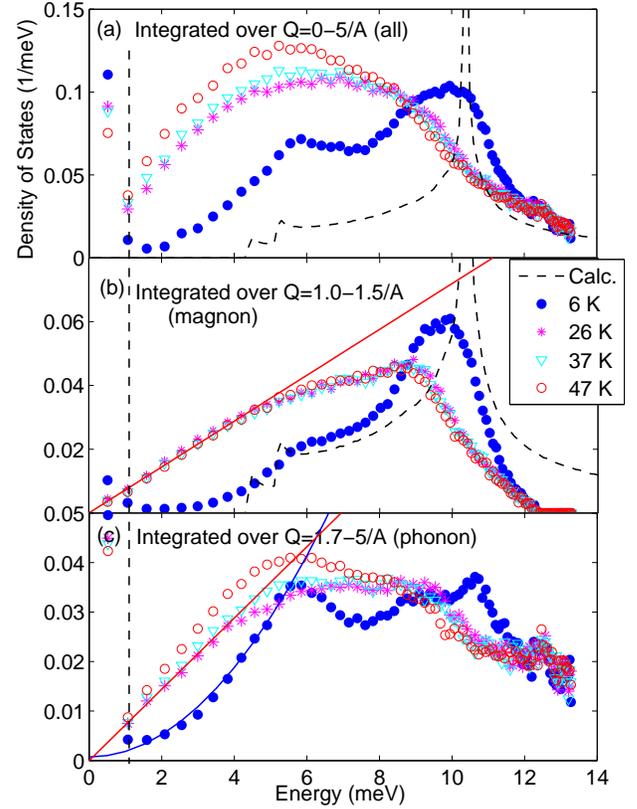}
\caption{\label{fig:ZZ} The density of states of magnon and phonons in s-O$_2$ (a) total excitations (b) magnetic component (c) phonon component.}
\end{figure}

We may gain insights into how the dynamics evolve by looking into the density of states. Dividing the dynamic structure function $S(Q,\omega)$ by the trivial temperature dependent factors, including the Debye-Waller factor $e^{-2W}$ and the phonon (or magnon) occupation number $n(\omega,T)$, and integrating over the appropriate $Q$ range allows us to extract the density of states of excitation modes of interest,
\begin{equation} 
%Z(\omega)=\sum_{Q} S(Q,\omega)=\int_{Q=\frac{1}{2}Q_{free}}^{\infty} S(Q,\omega) dQ 
Z(\omega)=\int_{Q_{min}}^{Q_{max}} \frac{\omega}{\hbar^2Q^2/2m}\frac{e^{2W}}{(n(\omega,T)+1)}S(Q,\omega) dQ,
\end{equation} 
where $[Q_{min},Q_{max}]=[1.0,1.5]\mbox{\AA}^{-1}$ for magnetic excitations and $[Q_{min},Q_{max}]=[1.7,5.0]\mbox{\AA}^{-1}$ for phonons. The resulting density of states are plotted in Fig.~\ref{fig:ZZ}.
The phonon density of states in $\alpha$-O$_2$ can be fitted by a quadratic function $\omega^2$, as expected, and the magnon density of states in $\alpha$-O$_2$ follows the theoretical predictions as discussed in sec.~\ref{sec:alpha}. For $\beta$-O$_2$, it is interesting to note that both nuclear and magnetic density of states can be fitted by a linear function in $\omega$ with the same slope, indicating that the fluctuations are correlated and might be due to the same cause. 
The density of states for $\gamma$-O$_2$ at high $Q$ is larger than that of $\beta$-O$_2$ as a result of the additional orbital rotational diffusion~\cite{Fernandez08}. 

\section{UCN cross sections}

%\begin{figure}
%\centering
%\includegraphics[width=3.5 in]{figures/SS_all.eps}
%\caption{UCN upscattering and S(Q,$\omega$)}
%\label{fig:UCN}
%\end{figure}

\begin{figure}
\centering
(a)\includegraphics[width=3.3 in]{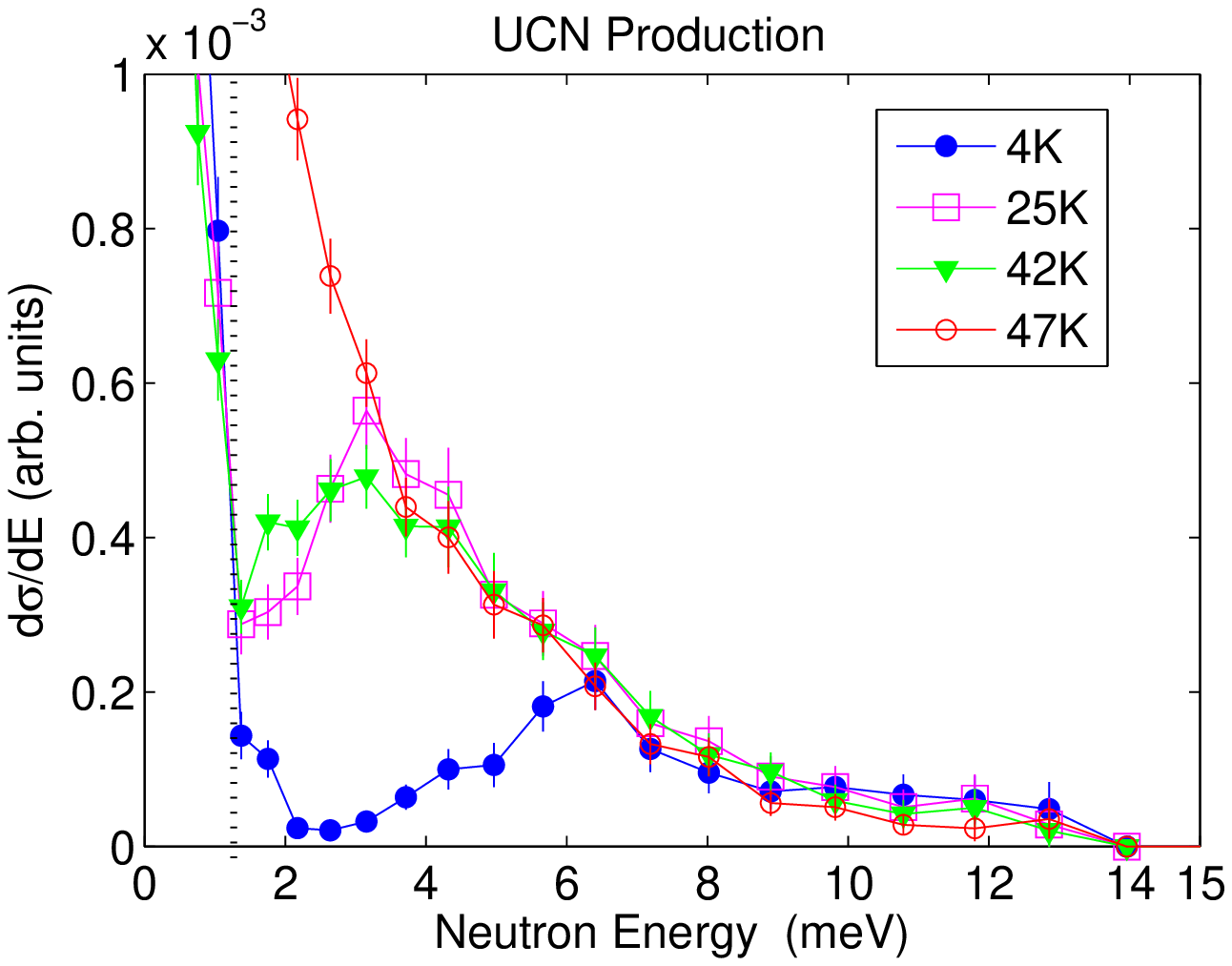}
(b)\includegraphics[width=3.3 in]{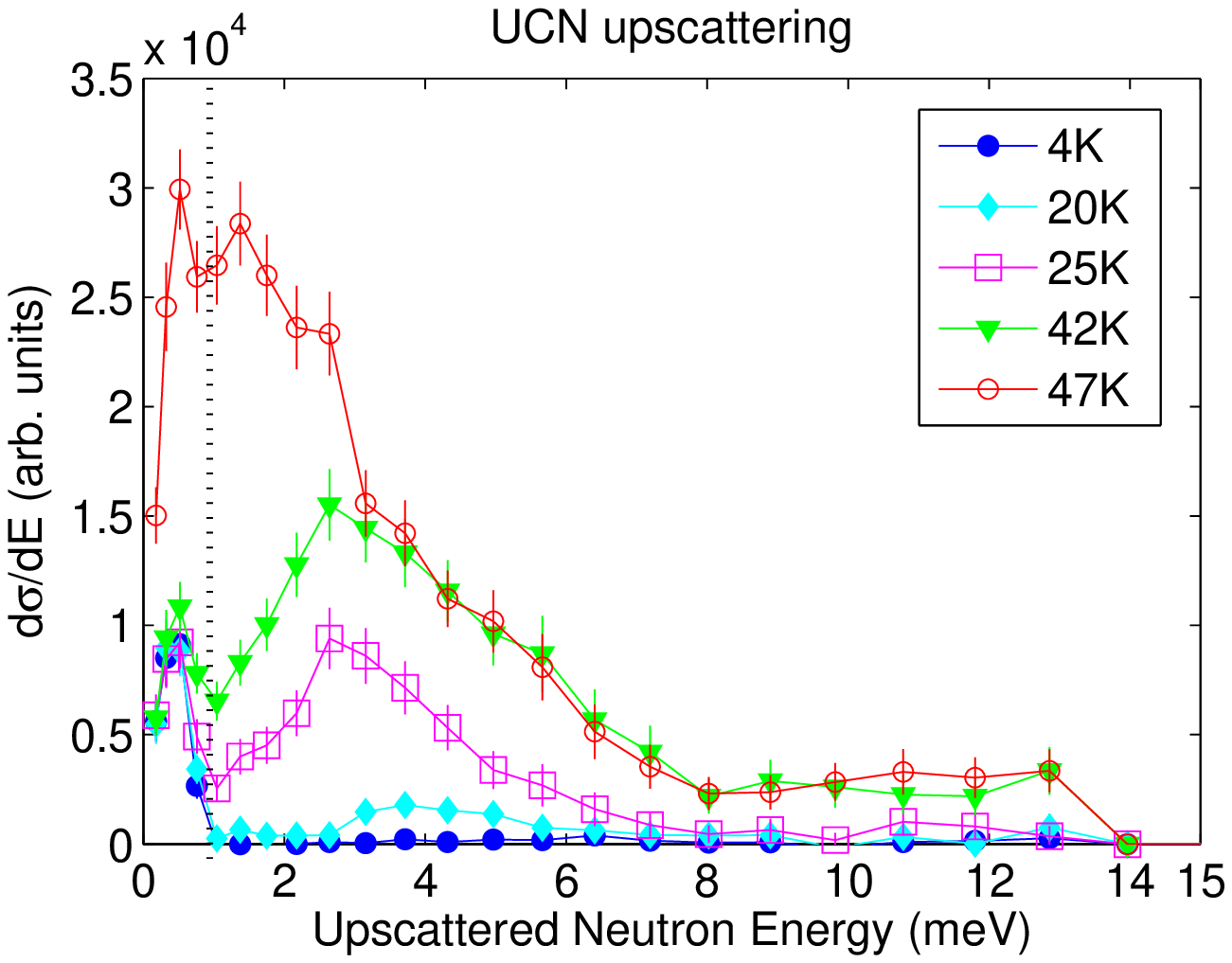}
\caption{(a) Differential cross sections for UCN production in s-O$_2$. (b) UCN upscattering cross sections in s-O$_2$. Note that in both plots the low E data ($<$ 1~meV) is contaminated by the finite width of elastic scattering.}
\label{fig:UCNprod}
\end{figure}

%\begin{figure}
%\centering
%\includegraphics[width=3.3 in]{figures/UCNup_summary2.eps}
%\caption{UCN upscattering cross sections in s-O$_2$. Note that the low E data ($<$ 1~meV) is contaminated by the finite width of elastic scattering.}
%\label{fig:UCNup}
%\end{figure}

The energy of UCN ($\sim$ 100 neV) is several orders of magnitudes smaller than the energy scale of cold neutrons ($\sim$ few meV).
For scatterings involving UCN production, because of the small momentum carried by UCN, $k_{ucn}$, the momentum transfer $Q$ is approximately equal to the initial momentum of the incident neutrons. 
Similarly, $Q$ is equal to the final momentum of the upscattered neutrons in UCN upscattering.
Therefore, the differential cross sections for UCN production (via downscattering) and UCN upscattering
can be attained by simply reading the value of $S(Q,\omega=E_{free})$ under the parabolic curve of free neutron dispersion, $E_{free}=\hbar^2Q^2/2m$, as illustrated by the solid red curve (for downscattering) and dashed curve (for upscattering) in Fig.~\ref{fig:SS_all}a, b, c, and Fig.~\ref{fig:dispersion}:  
\begin{eqnarray}
\frac{d\sigma^{down}}{dE} &=& 4\pi \frac{k_{ucn}}{Q} S\left(Q,\omega=\frac{\hbar^2Q^2}{2m_n}\right)_{\omega>0}, \label{Eq:UCNProduction} \\
\frac{d\sigma^{up}}{dE} &=& 4 \pi \frac{Q}{k_{ucn}} S\left(Q,\omega=-\frac{\hbar^2Q^2}{2m_n}\right)_{\omega<0}.
\label{Eq:UCNup}
\end{eqnarray}
At this stage, no effort was made to obtain the absolute cross section on the INS data, and the UCN cross sections are presented on a relative scale with an arbitrary normalization.
These temperature-dependent cross sections are then used as essential input parameters in the Monte-Carlo simulations to generate UCN yields that can be directly compared to the experimental UCN production data ~\cite{Liu2011}. If desired, the result of the comparison through a $\chi^2$ analysis yields the overall normalization required to set the absolute scale of the cross sections, but we omit this step here.

We plot the resulting 
(relative) differential cross sections for UCN production (Fig.~\ref{fig:UCNprod}a) and UCN upscattering (Fig.~\ref{fig:UCNprod}b) in s-O$_2$ in the $\alpha$ phase at 4~K, $\beta$ phase at 25~K and 42~K, and $\gamma$ phase at 47~K.
The differential cross section for UCN production can be integrated with the incident cold neutron spectrum to attain the UCN production rate specific to each experiment. Note that the production rate can be maximized by choosing an appropriate cold neutron spectrum that optimally overlaps with the differential cross section $\frac{d\sigma^{down}}{dE}$. As illustrated in Fig.~\ref{fig:UCNprod}a, the $\beta$-O$_2$ prefers a colder CN spectrum for maximum UCN production. The UCN upscattering, on the other hand, is not influenced by the CN spectrum, but instead is directly proportional to the number of magnons (or phonons) present in the solid and increases exponentially with increasing temperatures. To estimate the total UCN upscattering rate at different temperatures, we integrate Eq.~\ref{Eq:UCNup} over all final phase space.  

%\begin{equation}
%\sigma(T_{cn})=\int \frac{d\sigma(E_i\rightarrow E_{ucn})}{dE_i} \frac{E_i}{(k_{B}T_{cn})^2}e^{-E_i/k_BT_{cn}} dE_i
%\end{equation} 

\begin{figure}
\centering
\includegraphics[width=3.4 in]{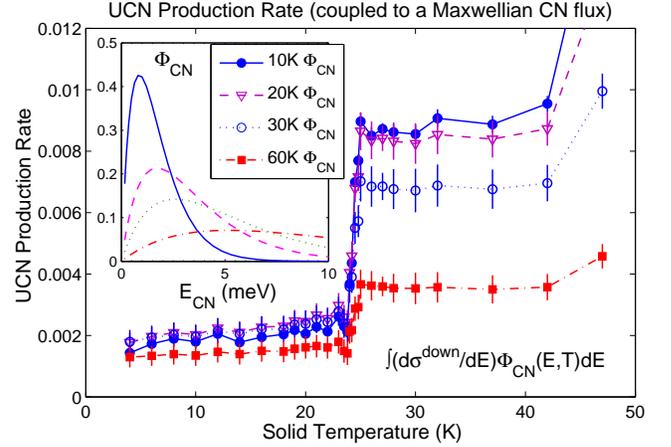}
\caption{Temperature-dependent UCN production rates in s-O$_2$, coupled to CN flux with a Maxwell-Boltzmann spectrum of 10~K, 20~K, 30~K and 60~K.}
\label{fig:CN_Tscan}
\end{figure}

A totally thermalized flux can be described by a Maxwell-Boltzmann distribution characterized by the mean temperature of the energy spectrum.
With a Maxwellian CN spectrum of temperature T$_{CN}$,  we can calculate the UCN production following the prescription described above to study the dependence of UCN production on the CN spectrum. The resulting temperature-dependent UCN production rates for s-O$_2$ coupled to CN flux of several T$_{CN}$ are shown in Fig.~\ref{fig:CN_Tscan}. Depending on the T$_{CN}$, the UCN production in $\alpha$-O$_2$ is 2$\sim$4 times smaller than that in the $\beta$-O$_2$. 
The dominant mechanism of UCN production in $\alpha$-O$_2$ is probably through phonon coupling around 6~meV, where the free neutron dispersion curve intersects with the phonon branch (as shown in Fig.~\ref{fig:SS_all}a). Unfortunately, due to the large spin anisotropy energy, the kinematical range where the magnon excitation becomes strong lies above and totally misses the free neutron dispersion curve. The magnon scattering, even though strong in the $\alpha$-O$_2$, is inaccessible to the inelastic scattering process that produces UCN through the superthermal principle. 
Above the phase transition, the spin dynamics changes into the highly unstable state with strong short-range correlations as a result of the geometrical frustration. The resulting dispersion-less magnetic excitations spread over $\omega$, extending down to $\omega=0$. These magnetic excitations overlap with the free neutron dispersion curve from $Q=1.0\sim 1.7 \mbox{\AA}^{-1}$, enhancing the UCN production for CN with energy ranging from $2\sim6$~meV (as illustrated in Fig.~\ref{fig:UCNprod}a). 
The UCN production in $\gamma$-O$_2$ is even larger, but the correspondingly large upscattering rate makes the UCN loss too big to be useful.

\section{Conclusion}
In this work, we update the seminal work of Stephens and coworkers~\cite{Stephens86} using the modern high resolution INS measurements.
Our INS data confirms the detailed prediction of the 2D AF magnon model, and we report new values of anisotropy energies that prefer the 2D Ising model over the XY model suggested by Stephens {\it et al}.  Our initial over-estimate of the UCN production in $\alpha$-O$_2$ is a result of the low values of spin anisotropy energy suggested by Stephens {\it et al}.
With the corrected anisotropy energies, we confirm that the UCN production in $\alpha$-O$_2$ is predominantly through phonon coupling, and thus suffers from the mass suppression factor because it is harder for a neutron to nuclear recoil the whole O$_2$ molecule that initiates the translational wave. In the $\beta$ phase, we have reported an unexpected large UCN production~\cite{Liu2011}. The INS measurements unfold the origin of the enhanced UCN production in $\beta$-O$_2$ as the unique dispersion-less magnetic excitations observed in other similar geometrically frustrated spin systems~\cite{Coldea03, Sato03, Stock09}. 
The spin dynamic in $\beta$-O$_2$ has been previously characterized using a phenomenological model of coupled paramagnets~\cite{Chahid93} on a three sublattice Yaffet-Kittel structure~\cite{Stephens86}, however, the ground state configuration must be highly unstable and the INS data shows many distinct features for the intriguing possibility for spinon excitations. The description of the detailed dynamics still awaits for better INS data using single crystal samples, which are challenging to prepare due to the thermal properties of oxygen.

We would like to thank the NIST neutron facility for providing the beam time and technical supports necessary to conduct this experiment.
We would like to thank Gerardo Ortiz for engaging discussions on the theoretical aspect of the work.
This work was supported by NSF grants 0457219, 0758018.

\bibliography{O2_NIST}

\begin{thebibliography}{26}
\expandafter\ifx\csname natexlab\endcsname\relax\def\natexlab#1{#1}\fi
\expandafter\ifx\csname bibnamefont\endcsname\relax
  \def\bibnamefont#1{#1}\fi
\expandafter\ifx\csname bibfnamefont\endcsname\relax
  \def\bibfnamefont#1{#1}\fi
\expandafter\ifx\csname citenamefont\endcsname\relax
  \def\citenamefont#1{#1}\fi
\expandafter\ifx\csname url\endcsname\relax
  \def\url#1{\texttt{#1}}\fi
\expandafter\ifx\csname urlprefix\endcsname\relax\def\urlprefix{URL }\fi
\providecommand{\bibinfo}[2]{#2}
\providecommand{\eprint}[2][]{\url{#2}}

\bibitem[{\citenamefont{Freiman and Jodl}(2004)}]{Freiman04}
\bibinfo{author}{\bibfnamefont{Y.~A.} \bibnamefont{Freiman}} \bibnamefont{and}
  \bibinfo{author}{\bibfnamefont{H.~J.} \bibnamefont{Jodl}},
  \bibinfo{journal}{Physics Reports} \textbf{\bibinfo{volume}{401}},
  \bibinfo{pages}{1} (\bibinfo{year}{2004}).

\bibitem[{\citenamefont{Lee et~al.}(1986)\citenamefont{Lee, Joannopoulos,
  Negele, and Landau}}]{Lee86}
\bibinfo{author}{\bibfnamefont{D.~H.} \bibnamefont{Lee}},
  \bibinfo{author}{\bibfnamefont{J.~D.} \bibnamefont{Joannopoulos}},
  \bibinfo{author}{\bibfnamefont{J.~W.} \bibnamefont{Negele}},
  \bibnamefont{and} \bibinfo{author}{\bibfnamefont{D.~P.}
  \bibnamefont{Landau}}, \bibinfo{journal}{Physical Review B}
  \textbf{\bibinfo{volume}{33}}, \bibinfo{pages}{450} (\bibinfo{year}{1986}).

\bibitem[{\citenamefont{Coldea et~al.}(2004)\citenamefont{Coldea, Tennant, and
  Tylczynski}}]{Coldea04}
\bibinfo{author}{\bibfnamefont{R.}~\bibnamefont{Coldea}},
  \bibinfo{author}{\bibfnamefont{D.~A.} \bibnamefont{Tennant}},
  \bibnamefont{and}
  \bibinfo{author}{\bibfnamefont{Z.}~\bibnamefont{Tylczynski}},
  \bibinfo{journal}{Journal of Magnetism and Magnetic Materials}
  \textbf{\bibinfo{volume}{272}}, \bibinfo{pages}{E649} (\bibinfo{year}{2004}).

\bibitem[{\citenamefont{Meier}(1985)}]{Meier85}
\bibinfo{author}{\bibfnamefont{R.~J.} \bibnamefont{Meier}},
  \bibinfo{journal}{Physics Letters A} \textbf{\bibinfo{volume}{112}},
  \bibinfo{pages}{341} (\bibinfo{year}{1985}) 

\bibitem[{\citenamefont{Jansen}(1986)}]{Jansen86}
\bibinfo{author}{\bibfnamefont{A.~P.~J.} \bibnamefont{Jansen}},
  \bibinfo{journal}{Physical Review B} \textbf{\bibinfo{volume}{33}},
  \bibinfo{pages}{6352} (\bibinfo{year}{1986}).

\bibitem[{\citenamefont{Jansen and Vanderavoird}(1987)}]{Jansen87a}
\bibinfo{author}{\bibfnamefont{A.~P.~J.} \bibnamefont{Jansen}}
  \bibnamefont{and}
  \bibinfo{author}{\bibfnamefont{A.}~\bibnamefont{Vanderavoird}},
  \bibinfo{journal}{Journal of Chemical Physics} \textbf{\bibinfo{volume}{86}},
  \bibinfo{pages}{3583} (\bibinfo{year}{1987}).

\bibitem[{\citenamefont{da~Silva and Falicov}(1995)}]{daSilva95}
\bibinfo{author}{\bibfnamefont{A.~J.~R.} \bibnamefont{da~Silva}}
  \bibnamefont{and} \bibinfo{author}{\bibfnamefont{L.~M.}
  \bibnamefont{Falicov}}, \bibinfo{journal}{Physical Review B}
  \textbf{\bibinfo{volume}{52}}, \bibinfo{pages}{2325} (\bibinfo{year}{1995}).

\bibitem[{\citenamefont{Lesar and Etters}(1988)}]{LeSar88}
\bibinfo{author}{\bibfnamefont{R.}~\bibnamefont{Lesar}} \bibnamefont{and}
  \bibinfo{author}{\bibfnamefont{R.~D.} \bibnamefont{Etters}},
  \bibinfo{journal}{Physical Review B} \textbf{\bibinfo{volume}{37}},
  \bibinfo{pages}{5364} (\bibinfo{year}{1988}). 

\bibitem[{\citenamefont{Stock et~al.}(2009)\citenamefont{Stock, Chapon,
  Adamopoulos, Lappas, Giot, Taylor, Green, Brown, and Radaelli}}]{Stock09}
\bibinfo{author}{\bibfnamefont{C.}~\bibnamefont{Stock}},
  \bibinfo{author}{\bibfnamefont{L.~C.} \bibnamefont{Chapon}},
  \bibinfo{author}{\bibfnamefont{O.}~\bibnamefont{Adamopoulos}},
  \bibinfo{author}{\bibfnamefont{A.}~\bibnamefont{Lappas}},
  \bibinfo{author}{\bibfnamefont{M.}~\bibnamefont{Giot}},
  \bibinfo{author}{\bibfnamefont{J.~W.} \bibnamefont{Taylor}},
  \bibinfo{author}{\bibfnamefont{M.~A.} \bibnamefont{Green}},
  \bibinfo{author}{\bibfnamefont{C.~M.} \bibnamefont{Brown}}, \bibnamefont{and}
  \bibinfo{author}{\bibfnamefont{P.~G.} \bibnamefont{Radaelli}},
  \bibinfo{journal}{Physical Review Letters} \textbf{\bibinfo{volume}{103}},
  (\bibinfo{year}{2009}). 

\bibitem[{\citenamefont{Sato et~al.}(2003)\citenamefont{Sato, Lee, Katsufuji,
  Masaki, Park, Copley, and Takagi}}]{Sato03}
\bibinfo{author}{\bibfnamefont{T.~J.} \bibnamefont{Sato}},
  \bibinfo{author}{\bibfnamefont{S.~H.} \bibnamefont{Lee}},
  \bibinfo{author}{\bibfnamefont{T.}~\bibnamefont{Katsufuji}},
  \bibinfo{author}{\bibfnamefont{M.}~\bibnamefont{Masaki}},
  \bibinfo{author}{\bibfnamefont{S.}~\bibnamefont{Park}},
  \bibinfo{author}{\bibfnamefont{J.~R.~D.} \bibnamefont{Copley}},
  \bibnamefont{and} \bibinfo{author}{\bibfnamefont{H.}~\bibnamefont{Takagi}},
  \bibinfo{journal}{Physical Review B} \textbf{\bibinfo{volume}{68}},
  \bibinfo{pages}{014432} (\bibinfo{year}{2003}). 

\bibitem[{\citenamefont{Lavelle et~al.}(2010)\citenamefont{Lavelle, Liu, Fox,
  Manus, McChesney, Salvat, Shin, Makela, Morris, Saunders
  et~al.}}]{Lavelle2010}
\bibinfo{author}{\bibfnamefont{C.~M.} \bibnamefont{Lavelle}},
  \bibinfo{author}{\bibfnamefont{C.~Y.} \bibnamefont{Liu}},
  \bibinfo{author}{\bibfnamefont{W.}~\bibnamefont{Fox}},
  \bibinfo{author}{\bibfnamefont{G.}~\bibnamefont{Manus}},
  \bibinfo{author}{\bibfnamefont{P.~M.} \bibnamefont{McChesney}},
  \bibinfo{author}{\bibfnamefont{D.~J.} \bibnamefont{Salvat}},
  \bibinfo{author}{\bibfnamefont{Y.}~\bibnamefont{Shin}},
  \bibinfo{author}{\bibfnamefont{M.}~\bibnamefont{Makela}},
  \bibinfo{author}{\bibfnamefont{C.}~\bibnamefont{Morris}},
  \bibinfo{author}{\bibfnamefont{A.}~\bibnamefont{Saunders}},
  \bibnamefont{et~al.}, \bibinfo{journal}{Physical Review C}
  \textbf{\bibinfo{volume}{82}}, \bibinfo{pages}{015502}
  (\bibinfo{year}{2010}). 

\bibitem[{\citenamefont{Copley and Cook}(2003)}]{Copley03}
\bibinfo{author}{\bibfnamefont{J.~R.~D.} \bibnamefont{Copley}}
  \bibnamefont{and} \bibinfo{author}{\bibfnamefont{J.~C.} \bibnamefont{Cook}},
  \bibinfo{journal}{Chemical Physics} \textbf{\bibinfo{volume}{292}},
  \bibinfo{pages}{477} (\bibinfo{year}{2003}). 

\bibitem[{\citenamefont{Liu et~al.}(2010)\citenamefont{Liu, Young, Lavelle, and
  Salvat}}]{Liu2010}
\bibinfo{author}{\bibfnamefont{C.-Y.} \bibnamefont{Liu}},
  \bibinfo{author}{\bibfnamefont{A.}~\bibnamefont{Young}},
  \bibinfo{author}{\bibfnamefont{C.}~\bibnamefont{Lavelle}}, \bibnamefont{and}
  \bibinfo{author}{\bibfnamefont{D.~J.} \bibnamefont{Salvat}}
  (\bibinfo{year}{2010}), \bibinfo{note}{arXiv:1005.1016v1 [nucl-th]}.

\bibitem[{\citenamefont{Stephens}(1985)}]{Stephens85}
\bibinfo{author}{\bibfnamefont{P.~W.} \bibnamefont{Stephens}},
  \bibinfo{journal}{Physical Review B} \textbf{\bibinfo{volume}{31}},
  \bibinfo{pages}{4491} (\bibinfo{year}{1985}). 

\bibitem[{\citenamefont{Dunstetter et~al.}(1996)\citenamefont{Dunstetter,
  Duparc, Plakhty, Schweizer, and Delapalme}}]{Dunstetter96}
\bibinfo{author}{\bibfnamefont{F.}~\bibnamefont{Dunstetter}},
  \bibinfo{author}{\bibfnamefont{O.~H.} \bibnamefont{Duparc}},
  \bibinfo{author}{\bibfnamefont{V.~P.} \bibnamefont{Plakhty}},
  \bibinfo{author}{\bibfnamefont{J.}~\bibnamefont{Schweizer}},
  \bibnamefont{and}
  \bibinfo{author}{\bibfnamefont{A.}~\bibnamefont{Delapalme}},
  \bibinfo{journal}{Low Temperature Physics} \textbf{\bibinfo{volume}{22}},
  \bibinfo{pages}{101} (\bibinfo{year}{1996}).

\bibitem[{\citenamefont{Collins}(1966)}]{Collins66}
\bibinfo{author}{\bibfnamefont{M.~F.} \bibnamefont{Collins}},
  \bibinfo{journal}{Proceedings of the Physical Society of London}
  \textbf{\bibinfo{volume}{89}}, \bibinfo{pages}{415} (\bibinfo{year}{1966}).

\bibitem[{\citenamefont{Stephens and Majkrzak}(1986)}]{Stephens86}
\bibinfo{author}{\bibfnamefont{P.~W.} \bibnamefont{Stephens}} \bibnamefont{and}
  \bibinfo{author}{\bibfnamefont{C.~F.} \bibnamefont{Majkrzak}},
  \bibinfo{journal}{Physical Review B} \textbf{\bibinfo{volume}{33}},
  \bibinfo{pages}{1} (\bibinfo{year}{1986}). 

\bibitem[{\citenamefont{Lindgard and et~al.}(1975)}]{Lindgard75}
\bibinfo{author}{\bibfnamefont{P.~A.} \bibnamefont{Lindgard}} \bibnamefont{and}
  \bibinfo{author}{\bibnamefont{et~al.}}, \bibinfo{journal}{Journal of Physics
  C: Solid State Physics} \textbf{\bibinfo{volume}{8}}, \bibinfo{pages}{1059}
  (\bibinfo{year}{1975}).

\bibitem[{\citenamefont{Liu and Young}(2004)}]{Liu2004}
\bibinfo{author}{\bibfnamefont{C.-Y.} \bibnamefont{Liu}} \bibnamefont{and}
  \bibinfo{author}{\bibfnamefont{A.}~\bibnamefont{Young}}
  (\bibinfo{year}{2004}), \bibinfo{note}{arXiv:nucl-th/0406004v1}.

\bibitem[{\citenamefont{Uyeda et~al.}(1985)\citenamefont{Uyeda, Sugiyama, and
  Date}}]{Uyeda85}
\bibinfo{author}{\bibfnamefont{C.}~\bibnamefont{Uyeda}},
  \bibinfo{author}{\bibfnamefont{K.}~\bibnamefont{Sugiyama}}, \bibnamefont{and}
  \bibinfo{author}{\bibfnamefont{M.}~\bibnamefont{Date}},
  \bibinfo{journal}{Journal of the Physical Society of Japan}
  \textbf{\bibinfo{volume}{54}}, \bibinfo{pages}{1107} (\bibinfo{year}{1985}).

\bibitem[{\citenamefont{Fernandez-Alonso
  et~al.}(2008)\citenamefont{Fernandez-Alonso, Bermejo, Bustinduy, Adams, and
  Taylor}}]{Fernandez08}
\bibinfo{author}{\bibfnamefont{F.}~\bibnamefont{Fernandez-Alonso}},
  \bibinfo{author}{\bibfnamefont{F.~J.} \bibnamefont{Bermejo}},
  \bibinfo{author}{\bibfnamefont{I.}~\bibnamefont{Bustinduy}},
  \bibinfo{author}{\bibfnamefont{M.~A.} \bibnamefont{Adams}}, \bibnamefont{and}
  \bibinfo{author}{\bibfnamefont{J.~W.} \bibnamefont{Taylor}},
  \bibinfo{journal}{Physical Review B} \textbf{\bibinfo{volume}{78}},
  (\bibinfo{year}{2008}). 

\bibitem[{\citenamefont{Rastelli et~al.}(1986)\citenamefont{Rastelli, Reatto,
  and Tassi}}]{Rastelli86}
\bibinfo{author}{\bibfnamefont{E.}~\bibnamefont{Rastelli}},
  \bibinfo{author}{\bibfnamefont{L.}~\bibnamefont{Reatto}}, \bibnamefont{and}
  \bibinfo{author}{\bibfnamefont{A.}~\bibnamefont{Tassi}},
  \bibinfo{journal}{Journal of Physics C-Solid State Physics}
  \textbf{\bibinfo{volume}{19}}, \bibinfo{pages}{L589} (\bibinfo{year}{1986}).

\bibitem[{\citenamefont{Rastelli and Tassi}(1988)}]{Rastelli88}
\bibinfo{author}{\bibfnamefont{E.}~\bibnamefont{Rastelli}} \bibnamefont{and}
  \bibinfo{author}{\bibfnamefont{A.}~\bibnamefont{Tassi}},
  \bibinfo{journal}{Journal of Physics C-Solid State Physics}
  \textbf{\bibinfo{volume}{21}}, \bibinfo{pages}{1003} (\bibinfo{year}{1988}).

\bibitem[{\citenamefont{Liu et~al.}(2011)\citenamefont{Liu, Lavelle, and
  Salvat}}]{Liu2011}
\bibinfo{author}{\bibfnamefont{C.}~\bibnamefont{Liu}},
  \bibinfo{author}{\bibfnamefont{C.}~\bibnamefont{Lavelle}}, \bibnamefont{and}
  \bibinfo{author}{\bibfnamefont{D.~J.} \bibnamefont{Salvat}}
  (\bibinfo{year}{2011}).

\bibitem[{\citenamefont{Coldea et~al.}(2003)\citenamefont{Coldea, Tennant, and
  Tylczynski}}]{Coldea03}
\bibinfo{author}{\bibfnamefont{R.}~\bibnamefont{Coldea}},
  \bibinfo{author}{\bibfnamefont{D.~A.} \bibnamefont{Tennant}},
  \bibnamefont{and}
  \bibinfo{author}{\bibfnamefont{Z.}~\bibnamefont{Tylczynski}},
  \bibinfo{journal}{Physical Review B} \textbf{\bibinfo{volume}{68}},
  (\bibinfo{year}{2003}). 

\bibitem[{\citenamefont{Chahid et~al.}(1993)\citenamefont{Chahid, Bermejo,
  Criado, Martinez, and Garciahernandez}}]{Chahid93}
\bibinfo{author}{\bibfnamefont{A.}~\bibnamefont{Chahid}},
  \bibinfo{author}{\bibfnamefont{F.~J.} \bibnamefont{Bermejo}},
  \bibinfo{author}{\bibfnamefont{A.}~\bibnamefont{Criado}},
  \bibinfo{author}{\bibfnamefont{J.~L.} \bibnamefont{Martinez}},
  \bibnamefont{and}
  \bibinfo{author}{\bibfnamefont{M.}~\bibnamefont{Garciahernandez}},
  \bibinfo{journal}{Journal of Physics-Condensed Matter}
  \textbf{\bibinfo{volume}{5}}, \bibinfo{pages}{6295} (\bibinfo{year}{1993}).

\end{thebibliography}

\end{document}